\PassOptionsToPackage{table,xcdraw}{xcolor}
\PassOptionsToPackage{hyphens}{url}
\documentclass[a4paper,fleqn]{cas-dc}
\usepackage[numbers,sort&compress]{natbib}
\usepackage{amsmath,amssymb,amsfonts}
\usepackage{amsthm}
\usepackage{algorithmic}
\usepackage{graphicx}
\usepackage{textcomp}
\usepackage{xspace}
\usepackage{listings}
\usepackage{enumitem}
\usepackage{booktabs}
\usepackage{caption}
\usepackage{wasysym}
\usepackage{xcolor}
\usepackage{multicol}
\usepackage{multirow}
\usepackage{soul}
\usepackage{float}
\usepackage{hyperref}
\hypersetup{pdfauthor={Daniel Gilkarov, Ran Dubin}}
\definecolor{accessblue}{cmyk}{1,0.3,0,0.2}
\newcommand{\fullcircle}{\CIRCLE}
\newcommand{\emptycircle}{\Circle}
\newcommand{\lefthalfcircle}{\LEFTcircle}

\def\BibTeX{{\rm B\kern-.05em{\sc i\kern-.025em b}\kern-.08em
    T\kern-.1667em\lower.7ex\hbox{E}\kern-.125emX}}
\begin{document}
\let\WriteBookmarks\relax
\def\floatpagepagefraction{1}
\def\textpagefraction{.001}

\shorttitle{NeuPerm: Disrupting Malware Extraction from Neural Network Parameters}
\shortauthors{D. Gilkarov and R. Dubin}

\title[mode = title]{\textbf{NeuPerm}: Disrupting the Extraction of Malware Hidden in \textbf{Neu}ral Network Parameters by Leveraging \textbf{Per}mutation Symmetry}

\author[1]{Daniel Gilkarov}[orcid=0009-0008-9274-802X]
\cormark[1]
\ead{daniel.gilkarov1@msmail.ariel.ac.il}

\author[1]{Ran Dubin}[orcid=0000-0002-2055-2211]
\ead{rand@ariel.ac.il}

\affiliation[1]{organization={Department of Computer and Software Engineering, Ariel University},
            city={Ariel},
            country={Israel}}

\cortext[1]{Corresponding author}

\begin{abstract}
Pretrained deep learning model sharing holds
tremendous value for researchers and
enterprises alike, but exposes end-users to
supply-chain attacks in which adversaries hide
self-executing malware inside neural network
parameters. In this work, we propose
\textbf{NeuPerm}, a zero-retraining,
zero-accuracy-cost randomized
parameter-reindexing transformation that
leverages \textbf{neu}ral network
\textbf{per}mutation symmetry to disrupt the
coordinate-dependent payload extraction used
by all currently-deployed NN-steganography
attacks, while provably preserving the model's
input-output function up to numerical
precision. We evaluate NeuPerm on nine CNN
architectures and two LLMs
(Llama-3.2-1B-Instruct and
Qwen2.5-1.5B-Instruct) against MaleficNet, a
state-of-the-art error-correcting
steganography attack, and a reimplementation
of EvilModel, a byte-substitution attack.
NeuPerm disrupts MaleficNet's
extraction, as designed,
on the evaluated architectures. NeuPerm further
disrupts the more fragile non error-correcting
EvilModel-style byte-substitution payload.
This comes with negligible impact on model
accuracy ($<$0.01\% on ImageNet, $\pm$0.5 F1
on SQuAD, $\pm$1\% accuracy on BoolQ, and
$<$0.001 perplexity on WikiText-2),
outperforming noise injection, pruning, and
quantization baselines, none of which disrupt
MaleficNet on the evaluated architectures. We complement the
empirical evaluation with a bound on the
success probability of coordinate-dependent
extraction under NeuPerm. NeuPerm defends
against currently-deployed, non-adaptive
attacks; a NeuPerm-aware adversary who embeds
the payload using a permutation-invariant
ordering (e.g., L1/L2 neuron norms) can
recover it by re-sorting after permutation. We
analyze such canonical-ordering attacks and
discuss future directions for defending
against them, such as generalizing NeuPerm to
additional weight-space symmetries. The source
code is available at
\url{https://github.com/danigil/NeuPerm.git}.
\end{abstract}

\begin{highlights}
\item NeuPerm permutes weights to disrupt coordinate-dependent stegomalware extraction
\item The transformation needs zero retraining and costs zero model accuracy
\item It disrupts MaleficNet and EvilModel-style attacks on nine CNNs and two LLMs
\item We bound the success probability of coordinate-dependent payload extraction
\item A permutation-invariant unshuffling adversary can still recover the payload
\end{highlights}


\maketitle

\section{Introduction}
Deep Learning (DL) neural network models are a major technological advancement that has been used and applied extensively in various fields: Computer Vision (CV) \cite{dl_cv_1,dl_cv_2,dl_cv_3}, Natural Language Processing (NLP) \cite{dl_nlp_1,dl_nlp_2,dl_nlp_3,dl_nlp_4}, in particular, the topic of Large Language Models (LLM) is drawing enormous attention \cite{dl_llm_1,dl_llm_2,dl_llm_3}, and many more \cite{dl_etc_1,dl_etc_2,dl_etc_3,dl_etc_4}. 
Neural network models, while very flexible and capable, require vast resources \cite{dl_ft} such as expertise, computation power, data, etc., to train models from scratch on a given task.
LLMs in particular are very useful, but they require a meticulous and resource-heavy training process. For example, training Meta's Llama1 65B LLM took 2048
A100 GPUs with 80GB of RAM, and the training process took approximately 21 days \cite{touvron2023llamaopenefficientfoundation}. 
The alternative is to use Pre-Trained Models (PTM) shared online on PTM hubs such as HuggingFace \cite{Huggingface}, and TensorFlow-Hub \cite{tensorflowhub}. The PTMs, when used correctly, serve as an enormous shortcut to applying a neural network model to some task \cite{dl_ft}. The necessity and value of PTM sharing are clear. However, PTM sharing introduces the potential for ML supply chain attacks, for example, attacks where attackers upload models infected with malware to the PTM sharing hubs \cite{stegonet,evilmodel1,evilmodel2,maleficnet1,maleficnet2}. The topic of ML supply chain attacks has been listed as a threat to ML security by different projects that map cyber threats: MITRE ATLAS (Adversarial Threat Landscape for Artificial-Intelligence Systems) \cite{mitre_atlas} indicates ML supply chain compromise as one of the threat vectors in Artificial-Intelligence (AI) systems, the OWASP top 10 for LLM applications 2025 also does so \cite{owasp_top_10_for_llms}. Securing the ML supply chain is paramount to ensuring end-user safety and continuous, unhindered collaboration on ML and DL.

Past research demonstrates how attackers can use data-hiding (steganography) techniques to hide malware inside neural network parameters. Song et al. \cite{dl_stego_1} initially introduced Least Significant Bit (LSB) substitution, sign-mapping (or sign-encoding), and correlated value encoding neural network steganography techniques that they used for data exfiltration attacks.
StegoNet \cite{stegonet}, a subsequent work, formulated the neural network steganography attack and researched 4 neural network steganography techniques - LSB substitution, sign-mapping, value-mapping, and resilience training. They design 2 trigger mechanisms, show the methods successfully evade detection by MetaDefender and traditional steganalysis and evaluate model accuracy before and after the attack, which causes up to 2\% test accuracy drops to CNNs in most cases.
EvilModel \cite{evilmodel1,evilmodel2} improved on StegoNet, they focused on byte substitution and aimed to reduce the performance drop the steganography causes. They claim almost half of the CNN models used could be attacked with up to 48.52\% Embedding Rate (ER) with no performance hit as opposed to only up to 15\% ER with StegoNet.
MaleficNet \cite{maleficnet1,maleficnet2} improved the previous neural network steganography works by focusing on increasing evasiveness, improving resilience to corruption of the payload by using error-correcting codes, and improving the trigger mechanism, which extracts and executes the malware upon loading of the model, instead of triggering based on some condition as in EvilModel \cite{evilmodel1,evilmodel2}.
After having a payload embedded in their parameters using neural network steganography, the resulting models are stegomalwares (malicious software that uses steganographic techniques) that can potentially cause damage.
These stegomalwares pose a security threat to the ML supply chain, as they can be easily propagated, do not degrade model performance \cite{evilmodel2,maleficnet1,maleficnet2}, and are shown to successfully avoid detection by common anti-virus software \cite{maleficnet2}, malware detection systems, and steganalysis methods \cite{dl_steganalysis_1}.
Moreover, as opposed to steganography carried out in traditional digital media types such as images \cite{img_stego_1}, audio \cite{audio_stego_1}, or text \cite{text_stego_1}, neural networks have a much greater capacity for hidden data, since they are much larger files. For illustration, JPG images usually weigh around 3 megabytes, and in contrast, neural network models sometimes weigh multiple gigabytes; the Llama3.3 70B LLM weighs approximately 140 gigabytes.

\begin{figure}
    \centering
    \includegraphics[width=1.0\linewidth]{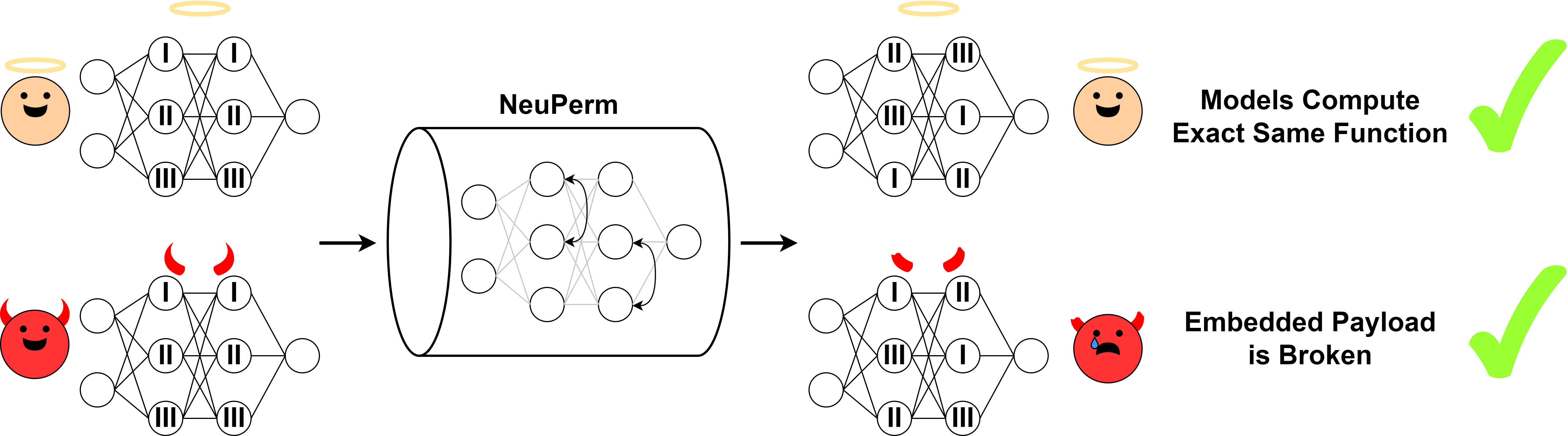}
    \caption{Illustration of \textbf{NeuPerm}. We permute the hidden units of the hidden layers of the neural network. The resulting models compute the same function as they did before the transformation. If the parameters contain an embedded payload, the attackers will not be able to extract it.}
    \label{fig:neuperm_illustration}
\end{figure}

In this work, we focus on disrupting the coordinate-dependent extraction of these neural network model stegomalwares.
Past efforts to combat neural network stegomalwares took 2 approaches: \textbf{(1)}: zero-trust mitigation - these works aim to damage a potential embedded payload by changing the parameters; adding random noise to the LSBs, and fine-tuning \cite{ran_dl_stego_cdr,amit_dl_stego_cdr} was shown to work on naive, non error-correcting neural network steganography like StegoNet and EvilModel \cite{stegonet,evilmodel1,evilmodel2}, but not on methods with error-correction such as MaleficNet \cite{maleficnet1,maleficnet2}; another method is compressing the weights using quantization \cite{ran_dl_stego_cdr}, which works even for the resilient error-correcting steganography introduced in MaleficNet \cite{maleficnet1,maleficnet2}, however, it lowers model performance \cite{ran_dl_stego_cdr} drastically as we demonstrate in experimental evaluation, and post-training quantization is highly specific \cite{krishnamoorthi2018quantizingdeepconvolutionalnetworks}.
\textbf{(2)}: steganography detection (steganalysis) -  these works aim to create techniques to classify samples as "innocent" (no payload) or "malicious", similar to steganalysis works in other domains \cite{img_stegan_1}. Zhao et al. \cite{dl_steganalysis_1} suggested steganalysis methods aimed to detect the methods introduced by Song et al. \cite{dl_stego_1}, which were proposed for data exfiltration attacks; while this is useful for these attacks, they might not generalize to unseen attacks like value-mapping(StegoNet) or MaleficNet \cite{stegonet,maleficnet1,maleficnet2}. Our previous work \cite{danigil_stega_1,danigil_stega_2} trained supervised classification models to detect LSB substitution steganography with an ER of up to 25\%, and we show trained methods generalize well to unseen attacks such as MaleficNet; however, the training process is complex, and it is fitted to a certain attack.
We note that these prior works
address \emph{steganalysis} - detecting whether a model
contains a hidden payload, whereas zero-trust
mitigation disrupts extraction independently of detection. The two approaches are complementary
and can be deployed together in a defense pipeline.
We present \textbf{NeuPerm}, a simple, yet effective zero-trust technique against this class of steganography that leverages the permutation symmetry of neural networks \cite{nn_param_perm}.
See Figure \ref{fig:neuperm_illustration}, our method has no measurable accuracy impact, it is simpler compared with model compression, and through evaluation, we conclude that it defeats resilient techniques such as MaleficNet on the evaluated architectures. To the best of our knowledge, NeuPerm is the first defense demonstrated to counter MaleficNet's spread-spectrum steganography, as designed, on the evaluated architectures; Section~\ref{sec:adaptive_unshuffle} characterizes the adaptive attacker that circumvents it.

While permutation symmetry is a well-known theoretical property, prior work has not applied it as a payload-disruption primitive. Unlike existing defenses that require retraining, architecture-specific tuning, or accept accuracy degradation, NeuPerm operates as a zero-retraining, zero-accuracy-cost post-processing step with a formal correctness guarantee, well suited to processing large volumes of untrusted models in ML supply chain security.

\noindent The key contributions of our research are:
\begin{itemize}
    \item We introduce \textbf{NeuPerm}: a
    zero-retraining, zero-accuracy-cost randomized
    parameter-reindexing transformation that disrupts
    coordinate-dependent neural network steganography
    via permutation symmetry.
    \begin{itemize}
        \item The first defense demonstrated to counter
        MaleficNet (as designed, on the evaluated
        architectures with $\geq$50\% measured permutable
        coverage) at zero accuracy cost, complementing
        existing detection-based defenses.
        \item Evaluated on 9~CNN architectures and two
        LLMs, covering modern designs including
        EfficientNet, MobileNet, ResNet, DenseNet, VGG,
        Llama-3.2-1B, and Qwen2.5-1.5B across
        3~datasets and 9~malware payloads, and compared
        against noise injection, pruning, and
        quantization baselines.
        \item Negligible performance impact: $<$0.01\%
        ImageNet accuracy change on CNNs, and on LLMs - three
        benchmarks spanning span extraction, binary
        classification, and language modeling ($\pm$0.5 F1
        on SQuAD, $\pm$1\% accuracy on BoolQ, $<$0.001
        perplexity on WikiText-2).
    \end{itemize}
    \item We analyze the security of NeuPerm against the
    coordinate-dependent extractor from both directions:
    a game-based analysis (in the definitional style of
    Katz and Lindell) that gives the exact success
    probability for attacks without error correction and
    unconditional bounds for those with it, and an
    adaptive permutation-invariant (unshuffling) attacker
    that recovers the payload, delimiting where the
    defense fails.
    \item We publish our source code at
    \url{https://github.com/danigil/NeuPerm.git}.
\end{itemize}

    

\section{Threat Model}
\label{sec:threat_model}
In this section, we present our threat model. Our threat model is similar to the one presented in past works such as StegoNet \cite{stegonet}, and MaleficNet \cite{maleficnet1}.
\subsection{System Entities}
\noindent The \textbf{End-User (Model Consumer)} may be any organization that ingests externally supplied model files, often using pretrained models from PTM hubs. They wish to leverage PTMs, mainly caring about the model performance.

\noindent The \textbf{Adversary} is an untrusted third-party that supplies malicious model files embedding executable or exfiltration payloads via steganographic techniques such as those demonstrated by StegoNet \cite{stegonet} or MaleficNet \cite{maleficnet1}. The adversary aims to launch an attack on the end-user via their malicious models, be it by causing RCE on the end-user's machine or by covertly delivering malicious payloads. 

\subsection{Adversary Capabilities and Assumptions}
Inspired by the StegoNet threat model \cite{stegonet} (Section 3), we adopt the following capabilities:

\begin{enumerate}
    \item \textbf{Supply-chain insertion.}  
          The adversary can publish or otherwise deliver a tampered model artifact to the consumer.
    \item \textbf{Model-only control.}  
          The adversary may arbitrarily modify \emph{static} contents of the model file (topology metadata, tensor values, \ldots) but \emph{cannot} directly execute code on the consumer’s host prior to model loading.
    \item \textbf{No direct network channel.}  
          Once the model is deployed within the consumer’s isolated environment, the adversary has no interactive connectivity to the host.
    \item \textbf{Trigger selection.}  
          The adversary can embed a \emph{logic bomb} that activates on specific inputs (e.g., logits-rank triggers) or upon deserialisation of the model (similar to MaleficNet \cite{maleficnet1}).
    \item \textbf{Coordinate-dependent extraction.}
      We assume the adversary's payload extraction procedure relies on fixed parameter coordinates (i.e., specific row and column indices in the weight matrices) to recover the embedded data. This assumption holds for all neural network steganography techniques published to date, including StegoNet \cite{stegonet}, EvilModel \cite{evilmodel1}, and MaleficNet \cite{maleficnet1}. We construct and evaluate a permutation-invariant extraction strategy in Section~\ref{sec:adaptive_unshuffle}.
\end{enumerate}

\subsection{Scope}
\label{sec:scope}
In this paper, we consider attacks that use the model file, specifically focusing on ones that use steganography. 
We separate the model-file attack surface into six
mechanisms that may be used separately or in tandem and discuss, for each, whether NeuPerm addresses it or not.
\begin{description}
\item[1. Unsafe deserialization:] arbitrary code execution or reconstruction of unsafe Python objects while the model file is parsed. \emph{Out of scope}: handled by the safe loader, which the recommended deployment applies before NeuPerm (Section~\ref{sec:deployment}).
\item[2. Parameter steganography:] a hidden payload embedded in tensor values. \emph{In scope}: this is our main target.
\item[3. Post-load execution:] running a payload after it has been extracted from the tensors. \emph{Out of scope}.
\item[4. Malicious model code:] executable logic in custom modules, repository scripts, or remote-code loading. \emph{Out of scope}.
\item[5. Behavioral backdoors:] trigger-based misclassification learned into the weights. \emph{Out of scope}.
\item[6. Model metadata attacks:] any payloads stored in the model metadata, such as chat templates in GGUF models. \emph{Out of scope}.
\end{description}

Safe loading and NeuPerm are complementary methods, because they act at different stages of
the threat. A safe loader neutralizes class~1 by
preventing code from executing while the file is parsed,
but it does not alter the parameter values, so a
steganographic payload (class~2) survives a safe load
intact. That residual payload is a covert channel with
more than one route to harm: it may be self-executing,
paired with a deserialization trigger that extracts and
runs it when the model is loaded, as in
MaleficNet~\cite{maleficnet1}, or it may simply carry
data that a separate mechanism recovers after delivery,
as the ZeusVM banking trojan did when it concealed its
configuration inside image files for later
retrieval~\cite{zeusvm_stego}. Safe loading addresses
only the first route, and only its execution step,
whereas NeuPerm disrupts the payload itself and therefore
complements safe loading against both.

Class~2's in-scope case is the extraction procedure that
assumption~(5) ascribes to every published technique.
Section~\ref{sec:adaptive_unshuffle} constructs the
adaptive, permutation-invariant extractor that defeats
it. Classes~4, 5, and 6 are out of scope and call
for complementary controls such as restricted-code
loading, sandboxed execution, content scanning, or
backdoor detection.

NeuPerm's transformation domain is the parameter tensors
of a recognized architecture; it reorders rows and
channels within them and does not read, rewrite, or
validate any other part of the artifact. By
assumption~(2), the adversary may place a payload outside
this domain: in the serialized-container or tensor-header
metadata (for example, the safetensors JSON header and
its unbounded \texttt{\_\_metadata\_\_} map), in
non-tensor files shipped alongside the weights, or in
parameters at non-permutable boundaries (quantified in
Section~\ref{sec:adaptive_nonperm}). Such payloads are
untouched by the permutation and require format-level
validation and content scanning at ingest.

NeuPerm is thus a method focused on steganographic
payloads: a randomized parameter-reindexing
transformation that disrupts existing
coordinate-dependent payload extraction.

\subsection{Deployment Model}
\label{sec:deployment}
NeuPerm is applied by the model consumer as a
post-download, pre-deployment transformation step.
Upon obtaining a model file from an untrusted source
(e.g., a PTM hub), the consumer safe-loads the file
into its parameter tensors (see below) and applies
NeuPerm to them with a freshly sampled random
permutation before the model enters production. The
permutation is generated locally and never shared, so
the adversary cannot observe or predict it. Since
NeuPerm operates as an in-place transformation of the
loaded parameters, with no retraining or fine-tuning
required, it can be integrated into existing
model ingestion pipelines as a single additional
processing step, analogous to checksum verification
or virus scanning.

\noindent\textbf{Safe loading in the deployment pipeline.}
Because assumption~(4) admits payloads that activate on
deserialization, the recommended pipeline safe-loads the
file before applying NeuPerm, which prevents such a
payload from executing while the tensors are parsed. Our approach to applying NeuPerm first loads the file for convenience, but it is possible to apply it to a dormant file before loading; we leave that to future work.
We recommend the \texttt{safetensors} format~\cite{safetensors}, whose Rust parser has no pickle path and
performs no code execution during load. For legacy
\texttt{.pt}/\texttt{.bin} checkpoints, the fallback is
\texttt{torch.load(weights\_only=True)} (the default
since PyTorch~2.6~\cite{pytorch260}), which restricts
unpickling to a safe allowlist rather than the general
pickle machinery; this narrows the attack surface but
does not fully close it.
The recommended pipeline
is: safe-load $\rightarrow$ permute (NeuPerm) $\rightarrow$
(optional steganalysis) $\rightarrow$ deploy.

\section{Neural Network Parameter Permutation}
\label{sec:neuperm_proof}
Neural network architectures often exhibit a property called parameter permutation symmetry \cite{nn_param_perm}, this property holds whenever neurons inside a hidden layer of the neural network can be permuted and the neural network still computes the same function. This can be expressed informally as:
\[\forall x, f(x)=f'(x)\]
where $f$ is some feedforward neural network, and $f'$ is the same neural network with some of its hidden layer units shuffled.
Past works thoroughly covered parameter symmetry, initially for MLPs, and channel permutation of CNNs \cite{nn_param_perm,nn_param_perm3,nn_param_perm4}, and later generalized to any arbitrary feedforward neural network architecture \cite{nn_param_perm2}.
See Section \ref{sec:perm_proof} for our proofs of the parameter permutation symmetry of select architectures.

\begin{figure}
    \centering
    \includegraphics[width=0.8\linewidth]{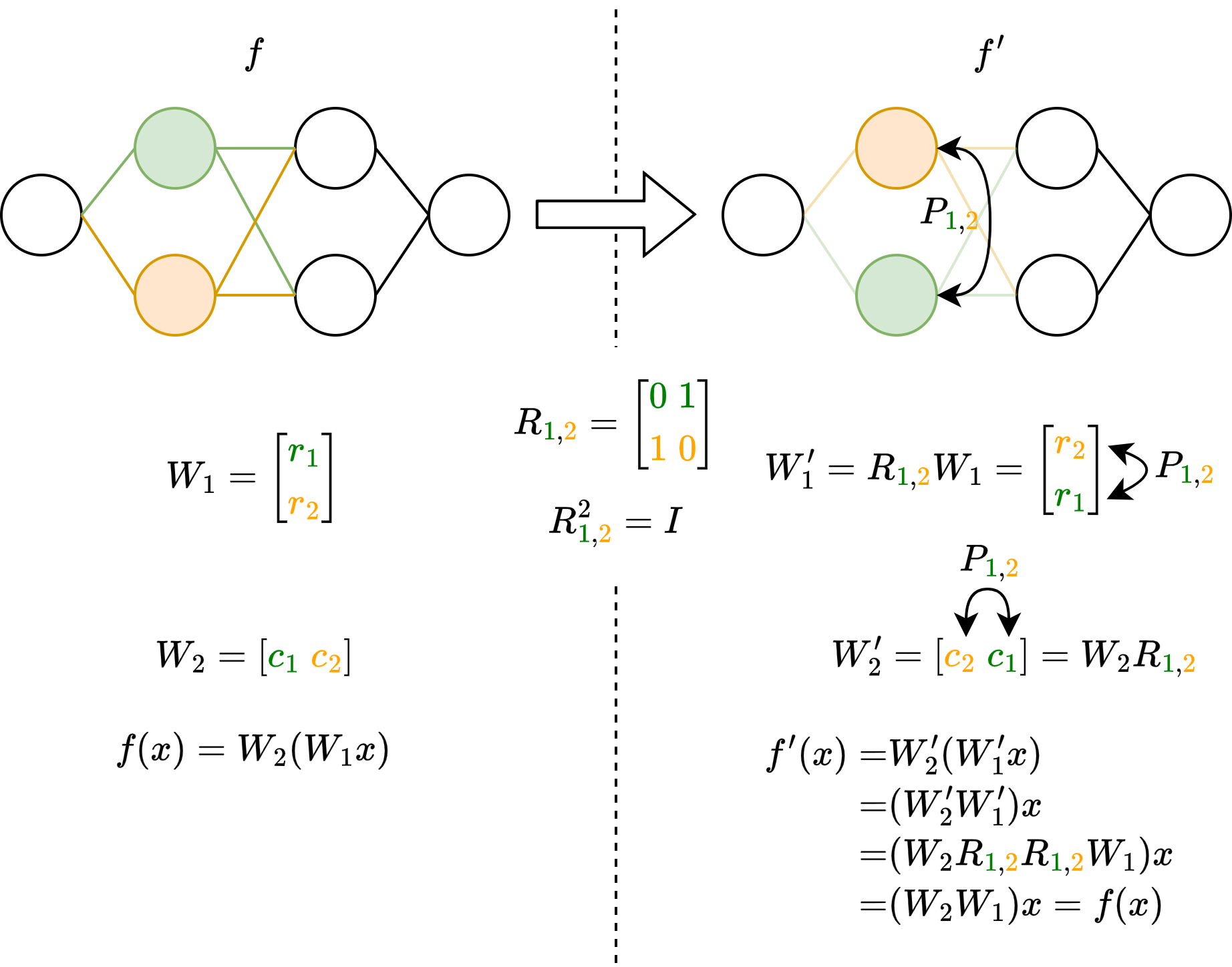}
    \caption{Multi-Layer Perceptron (MLP) parameter permutation symmetry. Swapping units in a hidden layer does not change the computation; however, it does change the parameter matrices.}
    \label{fig:mlp}
\end{figure}

\subsection{NeuPerm Implementation}
\label{sec:neuperm_implement}
We now describe how we apply NeuPerm in practice. The core principle is the same across all layer types: whenever a hidden layer's output units (neurons, or channels) can be reordered without affecting downstream computation, NeuPerm applies a random permutation to those units and propagates the reordering to the adjacent layers that consume them. This preserves the network's input-output function while shuffling the parameter coordinates that a steganographic payload depends on. Formal proofs that each building block admits this symmetry are provided in Appendix~\ref{sec:perm_proof}; here we describe the practical implementation for each block type.

Let us define some terms for convenience.
We use abbreviations to refer to commonly used neural network components: MLP/Fully Connected (FC), CNN/CONV (can be a 1D or 2D convolutional block), Batch Normalization (BN), Layer Normalization (LN), Average/Max Pooling (AVGP/MAXP), Embedding (EMB).
In the following sections, we express a sequential series of neural network blocks like so: BLOCK1-BLOCK2-$\cdots$; for example, CONV-BN-CONV means a convolutional block, followed by batch normalization, and finally, another convolutional block.

\subsubsection{FC-FC} See Figure \ref{fig:mlp}. Consider two consecutive fully connected layers with parameter matrices $W_1 \in \mathbb{R}^{n \times m}, b_1 \in \mathbb{R}^{n}$ (first layer, $n$ output units) and $W_2, b_2$ (second layer). Each row of $W_1$ and corresponding entry of $b_1$ computes one unit's pre-activation; permuting these rows simply reorders which unit occupies which index. Because the second layer reads from the first layer's output, we must apply the same reordering to the columns of $W_2$ so that each column still connects to the correct (now relocated) unit. As proven in Appendix~\ref{sec:perm_proof} (Claim~1.2), this joint permutation leaves the composition $W_2 \, \sigma(W_1 x + b_1) + b_2$ unchanged.

Pseudo-code:
Given a random permutation $PI$ of $[0, 1, \ldots, n{-}1]$:
\begin{lstlisting}
W1, b1 = W1[PI], b1[PI]
W2 = W2[:, PI]
\end{lstlisting}


\subsubsection{CONV-CONV/CONV-BN-CONV} Convolutional layers are analogous to FC layers with spatial weight sharing. In PyTorch/TensorFlow, a Conv2D layer with $n$ output channels and $m$ input channels using $k \times k$ filters stores its weights as a tensor $W \in \mathbb{R}^{n \times m \times k \times k}$. The first axis indexes output channels (analogous to rows in an FC layer), and the second axis indexes input channels (analogous to columns). The same permutation logic from the FC-FC case therefore applies: permuting the first axis of the first CONV reorders its output channels, and the matching permutation on the second axis of the next CONV ensures that each filter still reads from the correct (now relocated) input channel.

Pseudo-code:
Given permutation indexes $PI$:
\begin{lstlisting}
W1, b1 = W1[PI], b1[PI]
W2 = W2[:, PI, ...]
\end{lstlisting}
If there is an affine Batch Normalization (BN) layer between the two CONVs, its per-channel parameters ($\gamma$, $\beta$, running mean, running variance) must also be permuted along the channel axis to stay synchronized with the reordered channels. This is straightforward since BN operates independently per channel.


\subsubsection{MBConv (depthwise-separable / inverted-residual)} The MBConv block is the basic building block of many modern efficient CNNs (e.g., MobileNet, EfficientNet). An MBConv block expands the channel count inside the block, applies a depthwise convolution (optionally followed by a Squeeze-and-Excitation, SE, gate) on the expanded dimension, then projects back down, adding the result to the block input through a residual. Unlike a pointwise CONV, a depthwise convolution has a per-channel (diagonal) weight tensor of shape $(c,1,k,k)$, so it cannot absorb a permutation applied on only one side. The expanded dimension is, however, read and written by a whole chain of layers: the expansion CONV and its BN, the depthwise CONV (channel axis and BN), the optional SE FC pair, and the projection CONV input. Applying one shared permutation $PI$ to every layer that touches this dimension keeps them synchronized, because each producer and consumer of a channel is relabeled identically and the diagonal depthwise map simply follows the same reindexing. The block's external input and output orderings are never touched, so the residual connection is preserved. As proven in Appendix~\ref{sec:perm_proof}, functional equivalence then follows from the FC-FC and CONV claims applied around this ring.

Pseudo-code:
Given $PI$ a random permutation of size $c$ (the expanded width):
\begin{lstlisting}
# expansion CONV out-channels + BN
Wexp, Bexp = Wexp[PI], Bexp[PI]
# depthwise CONV (diagonal) + BN
Wdw, Bdw = Wdw[PI], Bdw[PI]
# SE gate FC1 out / FC2 in (if present)
Wse1, Bse1 = Wse1[PI], Bse1[PI]
Wse2 = Wse2[:, PI, ...]
# projection CONV in-channels
Wproj = Wproj[:, PI, ...]
\end{lstlisting}

\subsubsection{ATTN} Multi-Head Self-Attention (MHSA) computes each head independently and concatenates the results before a final linear projection (see Appendix~\ref{sec:perm_proof} for the formal computation). Because each head operates on its own slice of the representation, the heads can be reordered without affecting the block's output, provided the output projection matrix $W^O$ is adjusted to read from the reordered slices. This is the attention-level analogue of channel permutation in CNNs: heads play the role of channels, and the projection matrix plays the role of the next layer's weight matrix. We prove in Appendix~\ref{sec:perm_proof} that this head-level swap leaves the MultiHead computation unchanged.

Within each head, the internal units (rows of $W^Q$, $W^K$, $W^V$) can also be permuted, analogously to units in an FC layer, since they are consumed by the softmax-weighted combination and projected out through $W^O$.

Modern LLMs typically use Grouped Query Attention (GQA) \cite{dl_llm_2}, where $h_q$ query heads are divided into $G$ groups sharing $h_{kv}$ key-value heads. We permute at the key-value head granularity. Given an ATTN block with embedding size $d$: the query, key, value, and projection parameter matrices are $W_q \in \mathbb{R}^{h_{kv} \times G \times d}$, $W_o \in \mathbb{R}^{d \times G \times h_{kv}}$, $W_k, W_v \in \mathbb{R}^{h_{kv} \times d}$.

Pseudo-code:
Given $PI$, a random permutation of size $h_{kv}$:
\begin{lstlisting}
W_q, W_o = W_q[PI], W_o[..., PI]
W_k, W_v = W_k[PI], W_v[PI]
\end{lstlisting}
This permutes all $G$ query heads that share a given key-value head together, preserving the group structure.



\subsection{NeuPerm on Selected Architectures}
In this section, we detail how we apply NeuPerm to specific, well-known state-of-the-art neural network architectures. Section \ref{sec:neuperm_implement} describes how we apply NeuPerm to individual neural network layers, and now we specify how we choose to apply it to a full neural network.
We dub the strategies described here as "full NeuPerm". For evaluation, we also use "partial" versions of these strategies to analyze how effectiveness changes as NeuPerm permutes an increasing fraction of the model.
\subsubsection{Well-Known CNNs}
In this section, we detail how we apply NeuPerm to well-known CNNs. We use DenseNet, EfficientNet, MobileNet, ResNet, and VGG, which we consider representative of common CNN architectures.
Based on the theory \cite{nn_param_perm,nn_param_perm2,nn_param_perm3}, we can apply neural permutations in the following commonly used neural network blocks:
\begin{enumerate}
    \item FC-FC
    \item CONV-CONV
    \item CONV-BN-CONV
\end{enumerate}
This is enough to cover a significant portion of CNN architectures.
Note: Residual connections constrain permutation 
symmetry at specific layer boundaries: a CONV-CONV 
sequence with a residual connection in the middle 
does not admit permutation. However, this does not 
render NeuPerm inapplicable to residual architectures. 
As we detail below, NeuPerm permutes the inner 
layer sequences \emph{within} residual blocks 
(e.g., the CONV-BN-CONV sequences inside each 
Dense Layer or Residual Layer), achieving 76-88\% 
parameter coverage on DenseNet121 and 
ResNet50/101 (see Table~\ref{tab:neuperm_cnns}).

\noindent \textbf{DenseNet121}: DenseNet121 is structured like so: CONV-MAXP-DB1-T1-DB2-T2-DB3-T3-DB4-CLS.
Dense Blocks (DB) have nested Dense Layers (DL) with a CONV-BN-CONV sequence.
DLs within DBs are interconnected residually, which leads us to focus on just permuting the inner CONV-BN-CONV sequences within DLs.
Transition (T) blocks only have a CONV-AVGP sequence, which is not suited for permutation.
Likewise, the Classification (CLS) block is just a GLOBP-FC sequence.

\noindent \textbf{EfficientNet-B0/B4}: EfficientNet is a stack of MBConv blocks (each with an SE gate) between a CONV-BN stem and a CONV-BN-FC head. We permute the expanded dimension of every MBConv block via the shared ring described above, together with the permutable stem and head CONV-BN sequences. This changes $67\%$ (B0) and $86\%$ (B4) of parameters (Table~\ref{tab:neuperm_cnns}).

\noindent \textbf{MobileNet-V2/V3 (Small)}: The MobileNet family is likewise built from MBConv blocks. MobileNet-V2 has no SE gates; we permute the expanded dimension of every inverted-residual block via the ring, together with the stem and feature-to-head CONV-BN-CONV transitions, changing $51\%$ of parameters. MobileNet-V3 (Small) additionally contains SE subblocks, whose internal FC-FC pairs we permute, along with the final FC-FC classifier and the permutable stem transition, changing $81\%$ of parameters (Table~\ref{tab:neuperm_cnns}).

\noindent \textbf{ResNet50/101}: ResNet50 and ResNet101 are structured like so: CONV-MAXP-RB1-RB2-RB3-RB4-AVGP-FC.
They have a similar structure to the DenseNet121 network.
Residual Blocks (RB) have nested Residual Layers (RL) with a CONV-BN-CONV-BN-CONV sequence. RLs within RBs are interconnected residually.
Similar to DenseNet121, we choose to permute the inner CONV-BN-CONV-BN-CONV sequences.

\noindent \textbf{VGG11/16}: The VGG family of CNNs comprises just CONV-CONV sequences and two FC-FC sequences, without residual connections. We permute all of these sequences.

\noindent See Table \ref{tab:neuperm_cnns} for a summary of the percentage of parameters changed using the described strategies.

\addtolength{\tabcolsep}{3pt}

\begin{table}[]
\centering
\caption{Summary of the fraction of parameters changed by the full NeuPerm strategy for each evaluated architecture (CNNs and LLMs), over each model's learnable parameters with the tied LLM embedding counted once.}
\label{tab:neuperm_cnns}
\resizebox{\columnwidth}{!}{%

\begin{tabular}{llll}
\toprule
Model           & \#Parameters    & \begin{tabular}[c]{@{}l@{}}\#Changed\\ Parameters\end{tabular} & \begin{tabular}[c]{@{}l@{}}Fraction\\ of Changed\\ Parameters\end{tabular} \\ \midrule
\rowcolor[HTML]{EFEFEF} DenseNet121 & 8M              & 6.1M                                                           & 77\%                                                                       \\
ResNet50        & 25.6M           & 20.7M                                                          & 81\%                                                                       \\
\rowcolor[HTML]{EFEFEF} ResNet101   & 44.5M           & 39.6M                                                          & 89\%                                                                       \\
VGG11           & 132.9M          & 132.9M                                                         & 100\%                                                                      \\
\rowcolor[HTML]{EFEFEF} VGG16       & 138.4M          & 138.3M                                                         & 100\%                                                                      \\
EfficientNet-B0 & 5.3M            & 3.6M                                                           & 68\%                                                                       \\
\rowcolor[HTML]{EFEFEF} EfficientNet-B4 & 19.3M       & 16.7M                                                          & 86\%                                                                       \\
MobileNetV2     & 3.5M            & 1.8M                                                           & 51\%                                                                       \\
\rowcolor[HTML]{EFEFEF} MobileNetV3 (S) & 2.5M        & 2.1M                                                           & 81\%                                                                       \\
Llama-3.2-1B    & 1.24B           & 0.96B                                                          & 78\%                                                                       \\
\rowcolor[HTML]{EFEFEF} Qwen2.5-1.5B & 1.54B  & 1.24B                                                          & 81\%                                                                       \\ \bottomrule
\end{tabular}%

}
\end{table}

\addtolength{\tabcolsep}{-3pt}

\subsubsection{Well-Known LLMs}
\label{sec:neuperm_llm}
Modern decoder-only transformer LLMs are built by stacking identical transformer blocks, each containing an attention sub-block and a feed-forward (MLP) sub-block, with Layer Normalization (LN) and residual connections between them. NeuPerm is applicable to these architectures because the two core sub-blocks: ATTN and MLP; both admit permutation symmetry as described in Sections above and proven in Appendix~\ref{sec:perm_proof}. Importantly, the residual connections in a transformer block do \emph{not} prevent permutation of the sub-block internals: each residual adds the sub-block's output to the block's input along the embedding dimension, while NeuPerm permutes the \emph{internal} units of the sub-block (attention heads in ATTN, hidden units in MLP), which are projected back to the embedding dimension before the residual addition. The embedding-dimension ordering is therefore unaffected, and the residual connection is preserved.

The LN layers that precede each sub-block operate element-wise along the embedding dimension and do not interact with the internal unit ordering of ATTN or MLP, so they require no modification.

The only layers we do \emph{not} permute are the input embedding (EMB) and the final output projection (FC), since these serve as the model's interface to the vocabulary and their row/column ordering carries semantic meaning (token IDs).

We use the Llama-3.2-1B and Qwen2.5-1.5B models in our experiments. Because NeuPerm's permutation acts only on the LN-ATTN-LN-MLP block structure shared by decoder-only transformers, both the permutable coverage and the exact function-preservation apply directly to any model with the same block pattern (e.g., Mistral, Gemma) without modification.

\noindent \textbf{Llama-3.2-1B}: The Llama-3.2-1B model is structured as: EMB-LB$_1$-LB$_2$-$\cdots$-LB$_{16}$-FC, where each Llama Block (LB) contains the sequence LN-ATTN-LN-MLP. Within each LB, we apply two independent permutations: (1)~head permutation in the GQA attention block (as described in the ATTN subsection above), which shuffles the $W_q$, $W_k$, $W_v$, and $W_o$ matrices; and (2)~hidden-unit permutation in the MLP block, which contains an FC-FC structure (gate projection, up projection, and down projection) where we permute the intermediate hidden dimension. Across all 16 blocks, this results in $\approx 78\%$ of parameters changed.

\noindent \textbf{Qwen2.5-1.5B}: The Qwen2.5-1.5B model uses the same EMB-LB$_1$-$\cdots$-LB$_{28}$-FC structure with the same LN-ATTN-LN-MLP block, but with 28 layers (vs.\ Llama's 16), 12 attention heads (vs.\ 32) and 2 GQA key-value heads (vs.\ 8), 1536-dim embedding (vs.\ 2048), and 8960-dim MLP hidden (vs.\ 8192). Crucially, Qwen2.5-1.5B has Q/K/V projection \emph{biases} that Llama lacks; our permutation code reorders these biases jointly with the corresponding projection-weight rows so the model's input-output function is preserved (we verified this empirically: Qwen2.5-1.5B-Instruct's SQuAD F1 changes by $-0.03$ on average after applying NeuPerm). Across all 28 blocks, this results in $\approx 81\%$ of parameters changed.



\section{Formal Security Analysis}
\label{sec:formal}

\newcommand{\Enc}{\mathsf{Enc}}       
\newcommand{\Dec}{\mathsf{Dec}}       
\newcommand{\E}{\mathbb{E}}           
\newtheorem{lemma}{Lemma}[section]

Published neural network steganography attacks read their payload from fixed parameter coordinates: Song et al.\ \cite{dl_stego_1} and StegoNet \cite{stegonet} embed by sign mapping, value mapping, and LSB substitution, and EvilModel \cite{evilmodel1,evilmodel2} by byte substitution. Most carry no error correction, so extraction requires the stegomodel's parameters to remain unchanged, which their authors state, and this is why simple countermeasures such as additive parameter noise, fine-tuning, and pruning defeat them \cite{amit_dl_stego_cdr,ran_dl_stego_cdr}. MaleficNet \cite{maleficnet1,maleficnet2} is the exception: it embeds with an error-correcting code, and its authors demonstrate the payload surviving additive noise, fine-tuning, and pruning. Quantization was suggested against it, but mild 8-bit post-training quantization is insufficient, as our evaluation shows (Section~\ref{sec:exp2}). MaleficNet is therefore the hard case, and both the analysis and the evaluation center on it.

Our game-based analysis treats the coordinate-dependent adversary, whose extraction reads parameter coordinates fixed before the defender samples its permutation, and models each permutable layer as an independent uniform permutation on its output axis. For attacks without error correction it gives the success probability exactly (Section~\ref{sec:fsa_noecc}, and empirically Section~\ref{sec:exp3}); for attacks with error correction it gives unconditional bounds whose correctable fraction is bounded by MaleficNet's measured code rate rather than set to an assumed constant (Section~\ref{thm:bound}). Those bounds address each payload row by its original index, whereas MaleficNet's own despreading searches over alignments, so MaleficNet is covered empirically (Section~\ref{sec:exp2}). An adaptive adversary who embeds against a permutation-invariant coordinate system falls outside this scope and is treated in Section~\ref{sec:adaptive_unshuffle}.

\subsection{Model and Notation}
\label{sec:notation}

Let \(f_{\theta}\) be a neural network with parameter set \(\theta\), composed of \(m\) layers \(\mathcal{L}=\{1,\dots,m\}\).  Let
\[
  \mathcal{L}_{\mathrm{perm}}
  \;\subseteq\;
  \mathcal{L}
\]
be the subset of layers that admit permutation symmetry (e.g.\ fully-connected, convolutional, or attention blocks).  For each \(l\in\mathcal{L}_{\mathrm{perm}}\), denote:
\begin{itemize}
  \item \(n_l\): the \emph{output dimension} of layer \(l\) (number of units or convolutional channels).
  \item \(\Theta^{(l)}\in\mathbb{R}^{n_l\times d_l}\): the weight matrix or reshaped tensor whose first axis of length \(n_l\) is permutable.
  \item \(S_{n_l}\): the symmetric group on \(n_l\) elements.
\end{itemize}
Note: For simplicity, we assume each layer has one weight matrix.
Let $L'$ be the number of bits the adversary wishes to embed in the model.
NeuPerm operates by sampling, \emph{independently} for each \(l\in\mathcal{L}_{\mathrm{perm}}\),
\[
  P_l \;\stackrel{\$}{\leftarrow}\; \mathrm{Uniform}\bigl(S_{n_l}\bigr),
\]
and applying it to the first axis of \(\Theta^{(l)}\).  Equivalently, if \(\theta\) collects all parameters, define
\[
  \mathcal{G} \;=\;\prod_{l\in\mathcal{L}_{\mathrm{perm}}} S_{n_l},
\qquad
  P \;=\;(P_l)_{l\in\mathcal{L}_{\mathrm{perm}}}\in\mathcal{G},
\]
and let \(\widehat\theta = P(\theta)\) be the permuted parameters.  By construction \(f_{\theta}=f_{\widehat\theta}\).

\paragraph{Unit of analysis: permutable rows.}
The permutation \(P_l\) acts on the first axis of \(\Theta^{(l)}\), i.e.\ on entire \emph{rows} of length \(d_l\) - output channels for convolutional layers, output units for fully-connected layers, KV-head groups for grouped-query attention, and hidden units for MLP blocks.  All payload bits embedded in a single such row therefore share a common survival event: if the row is fixed by \(P_l\) they jointly survive in their original coordinates; otherwise they are jointly displaced.  We accordingly reformulate the analysis at the level of rows rather than individual bits.  Let \(R_l\) denote the number of rows of layer \(l\) used by the payload and let
\[
  R \;:=\; \sum_{l\in\mathcal{L}_{\mathrm{perm}}} R_l
\]
be the total number of payload-occupied rows.  Within each layer the relevant random object is the row permutation \(P_l\sim\mathrm{Uniform}(S_{n_l})\); across layers the family \(\{P_l\}_{l\in\mathcal{L}_{\mathrm{perm}}}\) is mutually independent.

\subsection{Security Game}
\label{sec:game}

A security definition has two components: a security
guarantee, which fixes what counts as a successful
attack, and a threat model, which fixes the power the
adversary is assumed to have~\cite{katz2015introduction}.
A threat model fixes the adversary's abilities, not the
strategy it uses within them.  We instantiate that
template for payload extraction under a randomized
reindexing defense.  The template is standard; the
instantiation below is ours.

\paragraph{Security guarantee (the break).}
The adversary breaks NeuPerm on a given instance if the
receiving stage of the attack reconstructs the payload
from the published model, i.e.\ if \(\widetilde m = m\).

\paragraph{Threat model (the adversary's power).}
We consider adversaries \(\mathcal{A}=(\Enc,\Dec)\)
whose readout is \emph{coordinate-dependent} and
\emph{non-adaptive} and whose decoder observes only the
coordinate-committed erasure channel defined below, which
is threat-model assumption~(5) of
Section~\ref{sec:threat_model} stated formally.  The embedding stage fixes, before
the defender's permutation is sampled, an injective
\emph{readout schedule}
\[
  \sigma:\{1,\dots,L'\}\;\longrightarrow\;
  \bigl\{(l,i,j)\bigr\},
\]
assigning to each payload position one parameter
coordinate.

The receiving stage does not observe the published
parameters directly; it observes one
\emph{coordinate-committed symbol} per payload position.
For payload position \(t\), write \(\sigma(t)=(l,i,j)\)
and let \(\rho(t)=(l,i)\) denote the layer and row of that
coordinate.  The row \(\rho(t)\) is \emph{fixed} under
\(P\) iff \(P_l(i)=i\).  The receiving stage then
observes
\[
  y_t
  \;=\;
  \begin{cases}
    \widehat\theta[\sigma(t)],
      & \text{if } P_l(i)=i
        \text{ (row } \rho(t) \text{ fixed)},\\[2pt]
    \bot,
      & \text{otherwise (row displaced: an erasure),}
  \end{cases}
\]
and outputs
\[
  \widetilde m
  \;=\;
  \Dec\bigl(y_1,\dots,y_{L'}\bigr),
  \quad
  \Dec:(\mathbb{R}\cup\{\bot\})^{L'}\to\{0,1\}^{L'}.
\]
Within this channel \(\Dec\) may be any function
whatsoever and of unbounded computational power, in
particular any error-correcting decoder; the restriction
is on its \emph{access}, not on its power.

The readout is committed in both respects: the
schedule \(\sigma\) is fixed before \(P\) is drawn, and
\(\Dec\) sees a displaced row only as \(\bot\).

\paragraph{The game.}
\begin{enumerate}[label=\arabic*.,leftmargin=*]
  \item A payload \(m\in\{0,1\}^{L'}\) is drawn uniformly at random and given to the embedding stage only.
  \item The embedding stage chooses a clean model \(\theta\) and outputs a stegomodel \(\theta^\star=\Enc(\theta,m)\) together with the readout schedule \(\sigma\).
  \item The defender samples \(P\sim\mathrm{Uniform}(\mathcal{G})\), independently of \((\theta^\star,\sigma)\), and publishes \(\widehat\theta=P(\theta^\star)\).
  \item The receiving stage is given \(\sigma\), \(\Dec\), the coordinate-committed symbols \((y_1,\dots,y_{L'})\), and black-box access to \(f_{\widehat\theta}\), and outputs \(\widetilde m=\Dec\bigl(y_1,\dots,y_{L'}\bigr)\), with \(y_t\) the coordinate-committed symbol defined above.
\end{enumerate}
The adversary's extraction success is
\[
  S
  \;:=\;
  \Pr\bigl[\widetilde m = m\bigr],
\]
the probability taken over the payload, the defender's
draw of \(P\), and any coins the adversary uses.  The
two stages model the sender and the receiver of the
covert channel: the receiving stage is not given \(m\),
since an attack whose receiver already holds the payload
has no channel to protect.  The black-box oracle in
step~4 conveys nothing about \(P\), since permutation
symmetry gives \(f_{\widehat\theta}=f_{\theta^\star}\)
(Section~\ref{sec:notation}) and the adversary
constructed \(\theta^\star\) itself.

\paragraph{What the class excludes.}
A random permutation hides \emph{indices}, not
\emph{values}.  \(P\) reorders the rows of each permutable
layer while leaving each row's contents intact, so NeuPerm
is in effect a transposition cipher over weight rows.  A
weight row's own content is a permutation-invariant
signature, so the permutation is efficiently invertible
from the units themselves, the classical anagramming
attack on a transposition cipher.

The class therefore excludes any decoder that recovers
the original ordering from the permuted values, since
such a decoder needs exactly the displaced values the
channel withholds.  Section~\ref{sec:adaptive_unshuffle}
constructs this adversary, a compliant raster read of
every coordinate followed by a re-sort of each layer's
units by their \(L_1\) weight norm inside \(\Dec\), and
measures it recovering the payload on every evaluated
architecture at a bit-error rate of at most \(0.001\)
(Table~\ref{tab:unshuffle_attack}).  It lies outside the
class by the decoder's access, not by any adaptivity: its
schedule is fixed before \(P\) exactly as the game
requires.  The bounds of Sections~\ref{sec:fsa_noecc}
and~\ref{thm:bound} concern coordinate-dependent
extraction and do not imply security against a
content-aware, unshuffling readout.  The class covers
every neural network steganography technique published to
date~\cite{stegonet,evilmodel1,maleficnet1}, each of
which reads its payload back from fixed parameter
indices.

\subsection{Fixed-Point Probability per Row}
\label{sec:fixed-point}

A coordinate-dependent embedding places each payload bit in some weight coordinate \((l,i,j)\) with \(l\in\mathcal{L}_{\mathrm{perm}}\) and \(i\in\{1,\dots,n_l\}\), so under a uniform \(P_l\) its survival is governed by whether row \(i\) of layer \(l\) is fixed:

\begin{lemma}\label{lem:fixed-prob}
Let \(X_{l,i}\in\{0,1\}\) indicate that row \(i\) of layer \(l\) is fixed by \(P_l\), i.e.\ \(P_l(i)=i\).  Then \(\Pr[X_{l,i}=1]=1/n_l\), and across layers the families \(\{X_{l,\cdot}\}_{l\in\mathcal{L}_{\mathrm{perm}}}\) are mutually independent.  Moreover, for any specified set \(I\subseteq\{1,\dots,n_l\}\) of \(r\) rows of a single layer \(l\),
\[
  \Pr\!\Bigl[\bigwedge_{i\in I} X_{l,i}=1\Bigr]
  \;=\;\frac{(n_l-r)!}{n_l!}
  \;=\;\prod_{j=0}^{r-1}\frac{1}{n_l-j}
  \;\ge\;\Bigl(\frac{1}{n_l}\Bigr)^{r},
\]
so the joint fixed-row probability is at least the product \((1/n_l)^r\) of the \(r\) marginal fixed-point probabilities, with equality only when \(r=1\).
\end{lemma}

\begin{proof}
Uniform randomness of \(P_l\) gives \(\Pr[P_l(i)=i]=1/n_l\).  Independence across layers follows from the independent draws of \(\{P_l\}\).  Within a layer, exactly \((n_l-r)!\) of the \(n_l!\) permutations fix a specified \(r\)-set, so the joint probability equals \((n_l-r)!/n_l!=\prod_{j=0}^{r-1}1/(n_l-j)\ge\prod_{j=0}^{r-1}1/n_l=(1/n_l)^r\).
\end{proof}

\subsection{Attacks With No Error-Correction}
\label{sec:fsa_noecc}
We first analyze the adversary's success probability \(S\) for an attack with no error-correction (Song et al.\ \cite{dl_stego_1}, StegoNet \cite{stegonet}, EvilModel \cite{evilmodel1}).  Such an attack succeeds only if \emph{every} payload-occupied row is fixed by its layer permutation, since any displaced row corrupts all bits it carries.  Letting \(I_l\subseteq\{1,\dots,n_l\}\) denote the set of \(R_l\) payload-occupied rows of layer \(l\), the success event factorizes across layers, and Lemma~\ref{lem:fixed-prob} gives the \emph{exact} probability
\[
    S \;=\; \Pr\!\Bigl[\bigwedge_{l\in\mathcal{L}_{\mathrm{perm}}}\;\bigwedge_{i\in I_l} X_{l,i}=1\Bigr]
      \;=\; \prod_{l\in\mathcal{L}_{\mathrm{perm}}}\frac{(n_l-R_l)!}{n_l!},
\]
using cross-layer independence of \(\{P_l\}\) together with the within-layer fixed-set probability of Lemma~\ref{lem:fixed-prob}, with \(R=\sum_l R_l\).  No inequality is needed, so we report the exact product.

In the MaleficNet configurations this paper evaluates, dense spreading saturates all but the narrowest permutable layer(s): the measured aggregate occupancy \(R/\sum_l n_l\) ranges from \(0.9595\) to \(1.0000\) across the 48 evaluated configurations, so most layers reach \(R_l=n_l\) while the narrowest need not (VGG16's first convolutional layer, for instance, carries the payload in \(16\) of its \(64\) rows, per the per-layer geometry recorded in the accompanying artifact).  The exact product above is authoritative for whatever per-layer \(R_l\) obtains, and at the measured occupancy it is astronomically small: for the VGG16 running example carrying the Stuxnet payload it evaluates to approximately \(10^{-33{,}941}\).  In the fully-saturated special case \(R_l=n_l\) it simplifies to the idealization
\[
    S \;=\; \prod_{l\in\mathcal{L}_{\mathrm{perm}}}\frac{1}{n_l!},
\]
in which a single permutable layer of width \(n_l\) already contributes a factor \(1/n_l!\).  The guarantee requires only that the payload occupy rows lying within the permutable layers; rows in non-permutable layers fall outside the exact product and are treated separately (Section~\ref{sec:limitation}).  For any non-error-correcting attack meeting that condition, the exact product at the measured occupancy drives the extraction probability to effectively zero.

\subsection{Attacks With Error-Correction}
\label{thm:bound}
We now turn to the harder case of an error-resilient attack.  We bound an idealized fixed-coordinate readout that addresses each payload row by its original index and whose code corrects a \(\delta\)-fraction of row \emph{erasures}, with \(\delta\) taken from MaleficNet's LDPC parameters \cite{maleficnet1,maleficnet2}.  This abstraction and MaleficNet's alignment-search despreading are not comparable instance-wise (Section~\ref{sec:fsa_assumptions}; MaleficNet's own readout is evaluated empirically in Section~\ref{sec:exp2}).  Decoding succeeds only if at least \((1-\delta)R\) of the \(R\) payload-occupied rows survive their layer permutation, i.e.\ only if \(C\ge k\), where
\[
  C \;:=\;\sum_{l\in\mathcal{L}_{\mathrm{perm}}}\sum_{i\in I_l} X_{l,i},
  \qquad
  k \;:=\;\lceil (1-\delta)R\rceil .
\]

A payload row displaced by its layer permutation reaches \(\Dec\) as the erasure symbol \(\bot\) by the game's receiving stage (Section~\ref{sec:game}), so it carries none of that row's payload bits, and \(\widetilde m=m\) therefore requires \(C\ge k\):
\[
  S \;=\; \Pr\bigl[\widetilde m = m\bigr] \;\le\; \Pr[C\ge k].
\]

\paragraph{Expected survival.}
By linearity of expectation and Lemma~\ref{lem:fixed-prob},
\[
  \E[C] \;=\; \sum_{l\in\mathcal{L}_{\mathrm{perm}}}\frac{R_l}{n_l}.
\]
Each term is the fraction of layer \(l\)'s rows the payload occupies, so \(\E[C]\) is bounded by the number of permutable layers and never by \(R\).  In the evaluated regime \(R_l=n_l\) this collapses exactly to
\[
  \E[C] \;=\; m_{\mathrm{p}}, \qquad m_{\mathrm{p}}:=|\mathcal{L}_{\mathrm{perm}}|,
\]
the classical fact that a uniform permutation has exactly one fixed point in expectation, applied per permutable layer.  NeuPerm therefore leaves, in expectation, exactly one payload-carrying row per permutable layer (14 for VGG16 and 66 for ResNet101) against the \(k\) rows the code must recover, which are 8{,}930 and 12{,}195 respectively.  The gap between one surviving row and several thousand required rows is the entire mechanism by which NeuPerm disrupts a coordinate-dependent error-correcting code.

\paragraph{Markov bound (assumption-free).}
Because \(C\ge0\), Markov's inequality gives, with no assumption about the dependence among the fixed-point indicators,
\[
  \Pr[C\ge k] \;\le\; \frac{\E[C]}{(1-\delta)R} \;=\; \frac{\sum_l R_l/n_l}{(1-\delta)R}.
\]
This holds for any decoder and any payload placement; it is the conservative floor.

\paragraph{First-moment bound (cross-layer independence).}
A sharper bound follows from a first-moment argument over \(k\)-subsets~\cite{mitzenmacher2017probability}.  If \(C\ge k\) then some \(k\)-subset of the marked rows is entirely fixed, so \(\binom{C}{k}\ge1\) and \(\Pr[C\ge k]\le\E\binom{C}{k}\).  Expanding across independent layers,
\[
  \Pr[C\ge k] \;\le\; \E\binom{C}{k}
  \;=\!\!\sum_{\substack{(k_l):\,\sum_l k_l=k\\ 0\le k_l\le R_l}}\;\prod_{l}\binom{R_l}{k_l}\frac{(n_l-k_l)!}{n_l!},
\]
where each per-layer factor is the exact fixed-\(k_l\)-set probability of Lemma~\ref{lem:fixed-prob}.  In the evaluated regime \(R_l=n_l\), each layer's fixed-point count satisfies \(\E\binom{C_l}{j}=1/j!\), and the product form admits the closed upper bound
\[
  \Pr[C\ge k] \;\le\; \frac{m_{\mathrm{p}}^{k}}{k!}.
\]
We claim no tightness, because the final step sums over all compositions of \(k\) and the Markov step on \(\binom{C}{k}\) is already loose; the exactness that holds is on the first moment alone, \(\E\binom{C}{k}=m_{\mathrm{p}}^k/k!\) for \(k\le\min_l n_l\), while the probability it bounds is strictly smaller.  We keep the product form for computation and report \(m_{\mathrm{p}}^k/k!\) as its closed-form summary.

\paragraph{The correctable fraction \(\delta\).}
MaleficNet's decoder is belief propagation against an additive-Gaussian channel, under which a permuted row becomes noise rather than a declared erasure, and an LDPC ensemble's decoding threshold is a property of its degree distribution rather than a constant readable from the source; in its place we use the information-theoretic ceiling on \emph{guaranteed unique decoding}.  MaleficNet's code is \texttt{make\_ldpc}(12192,\,\(d_v{=}3\),\,\(d_c{=}12\)), an LDPC code of block length \(N=12192\) whose parity-check matrix has \(3{,}048\) rows of rank only \(3{,}046\), a rank deficiency of two, so that \(K=9146\) and \(N-K=3046\) and no decoder can uniquely decode once more than an \((N-K)/N=3046/12192=0.24984\) fraction of the rows per block is erased, giving \(\delta\le0.24984\).  This is the \emph{erasure} ceiling with the erased positions known; an adversary lacking those positions faces the smaller errors ceiling \((1-\rho)/2=0.12492\) for the code rate \(\rho=K/N=0.75016\).  Taking this ceiling makes both bounds \emph{decoder-independent}: they hold for any error-correcting decoder the adversary might substitute.

We evaluate both bounds at \(\delta=0.24984\) for every model in our evaluation, with \(k=\lceil(1-\delta)R\rceil\).  Every constant is measured rather than assumed: \(m_{\mathrm{p}}\), the number of permutable layers, and the per-layer \(n_l\) come from the permuters; \(R=\sum_l n_l\), the total width of the permutable layers, is used in place of the count of payload-occupied rows and justified by the measured payload occupancy of Section~\ref{sec:fsa_noecc}; and \(\delta\) is bounded by the measured code rate.

Three configurations bracket the range.  VGG16, the running example, has \(m_{\mathrm{p}}=14\) permutable layers, \(R=11{,}904\) payload-occupied rows, and \(k=8{,}930\) rows to recover, at which the assumption-free Markov floor is \(1.568\times10^{-3}\) and the first-moment bound is \(10^{-21{,}170}\).  DenseNet121 is the worst case on both axes, with the loosest Markov value and the tightest redundancy margin: at \(m_{\mathrm{p}}=58\), \(R=7{,}424\), and \(k=5{,}570\), its Markov floor is \(1.041\times10^{-2}\) and its first-moment bound \(10^{-8{,}625}\).  Qwen2.5-1.5B is the strongest on MaleficNet's dense-spreading attack, with \(m_{\mathrm{p}}=56\), \(R=250{,}936\), and \(k=188{,}244\), giving a Markov floor of \(2.975\times10^{-4}\) and a first-moment bound of \(10^{-582{,}098}\).

That ranking is specific to the dense-spreading payload MaleficNet deploys: because the bounds are worst-case over row selection (assumption~\ref{asm:rowselect}), an adversary could instead concentrate its payload in one of Qwen's \(28\) permutable layers of width \(n_l=2\) (the \texttt{self\_attn.q\_proj} projections), where a single row survives with probability \(\tfrac12\) and the Markov floor rises to at most \(1/((1-\delta)\cdot2)=0.667\).  The per-model guarantee is thus governed by the widths of the \emph{occupied} layers rather than by how much payload is embedded, which is the regime that remark~\ref{rem:concentration} states abstractly.  Across them the Markov floor (the assumption-free conservative statement) stays at most \(1.041\times10^{-2}\), and below \(3.3\times10^{-4}\) on the two LLMs, while the first-moment bound, requiring only cross-layer independence, is smaller by hundreds of orders of magnitude.  Of the two, the first-moment bound is the operative guarantee, and the quantitative claim against the coordinate-dependent adversary rests on it.  The uniformity and cross-layer independence that the first-moment bound requires are properties of NeuPerm's own sampling, which draws each \(P_l\) uniformly and independently as Section~\ref{sec:notation} specifies, so a correct deployment satisfies them by construction.  This negligibility does not follow from full permutable coverage alone: the closed first-moment bound \(m_{\mathrm{p}}^{k}/k!\) falls below one once the correction budget clears the redundancy margin \(k>e\,m_{\mathrm{p}}\), equivalently \(\rho\,(R/m_{\mathrm{p}})>e\) for the code rate \(\rho=0.75016\) and the average permutable-layer width \(R/m_{\mathrm{p}}\); the exact crossover \(k!>m_{\mathrm{p}}^{k}\) is 36 for VGG16 (\(m_{\mathrm{p}}=14\)) and ranges from 22 to 176 across the evaluated configurations, against \(k=8{,}930\) on VGG16.  A configuration satisfying every other hypothesis but not this one can leave the success probability of order one half.  Every evaluated configuration clears the threshold by at least 35-fold, tightest on DenseNet121; a coordinate-dependent error-correcting attack therefore has negligible success probability on each of them.

\subsection{Assumptions and Scope of the Bounds}
\label{sec:fsa_assumptions}
Both bounds rest on a small, explicit set of assumptions, which we state and verify in turn.
\begin{enumerate}[leftmargin=*]
  \item\label{asm:uniform} Each layer permutation \(P_l\) is drawn uniformly from the symmetric group \(S_{n_l}\).  This per-permutation uniformity fixes the marginal fixed-point probability at \(1/n_l\), and the exact product \(S\), the first-moment bound, and the Markov floor all depend on it.
  \item\label{asm:independent} The permutations are drawn independently across layers.  The exact product \(S\) and the evaluation of the first moment \(\E\binom{C}{k}\) as a product over layers require this independence, whose factorization fails without it.  The first-moment \emph{inequality} \(\Pr[C\ge k]\le\E\binom{C}{k}\) does not: it is a Markov step on \(\binom{C}{k}\ge1\).  Neither does the Markov floor, which follows from linearity of expectation alone and remains valid under permutations correlated across layers, provided each stays marginally uniform.
  \item\label{asm:rowselect} The adversary may mark any \(R_l\) distinct rows in each permutable layer; both bounds are worst-case over that selection and make no uniformity assumption about which rows are chosen.
  \item\label{asm:nonperm} \(R\) counts only the payload-occupied rows lying within the permutable layers \(\mathcal{L}_{\mathrm{perm}}\); rows in non-permutable layers survive with probability 1, lie outside both bounds, and are treated in Section~\ref{sec:adaptive_nonperm}.
  \item\label{asm:defender} The defender neither knows nor controls the embedding distribution; the permutation is sampled independently of it.
  \item\label{asm:integrality} The Markov floor requires only \(k>0\).  The first-moment bound additionally requires \(k\) to be a positive integer, since its derivation passes through ``\(C\ge k\) implies \(\binom{C}{k}\ge1\)'', meaningful only for integer \(k\); the definition \(k=\lceil(1-\delta)R\rceil\) supplies one throughout.
\end{enumerate}
Three further remarks fix the scope of the bounds and their sensitivity to how the payload is placed.
\begin{enumerate}[leftmargin=*, start=7]
  \item\label{rem:concentration} The first-moment bound \(m_{\mathrm{p}}^{k}/k!\), applied as an upper bound outside the \(R_l=n_l\) regime in which the first moment is exact, weakens as the payload concentrates and \(R\) falls: on two synthetic layers of width 512 it rises from \(10^{-526}\) at \(R=400\) to \(10^{-8.5}\) at \(R=20\).  The Markov floor is not monotone in \(R\).  Because \(\E[C]/((1-\delta)R)=\tfrac{1}{1-\delta}\cdot(\sum_l R_l/n_l)/(\sum_l R_l)\) is a weighted mean of the per-layer \(1/n_l\), it is invariant in \(R\) when all permutable layers share a common width (measured at \(0.00260\) for \(R=400,200,100,40,20\) alike) and otherwise moves with which layer the payload occupies: for widths \(n=[16,512]\) at fixed \(R=16\), it reads \(0.04296\) spread, \(0.08332\) concentrated in the narrow layer, and \(0.00260\) in the wide one.
  \item\label{rem:width} Each layer width \(n_l\) enters the product individually, with no reduction to a single \(n\): the collapsed single-layer form \(\binom{R}{k}(n-k)!/n!\) is invalid whenever \(k>n_{\min}\), which holds for most evaluated models because \(k\) is a fraction of the \emph{total} payload row count while \(n_{\min}\) is the width of the narrowest single layer.
  \item\label{rem:maleficnet} The bounds govern an idealized fixed-coordinate readout that addresses each row by its original index, so any displaced row counts as an erasure.  MaleficNet's spread-spectrum despreading is instead an alignment search that reconstructs from the \emph{location} of a correlation peak, so a permutation acting as a cyclic shift displaces every row, an erasure under our model, while merely relocating the real peak at full height, an instance where the modeled readout fails and the real one succeeds.  The two are therefore not comparable instance-wise, and no claim of dominance is made; Section~\ref{sec:exp2} covers MaleficNet empirically, demonstrating the despreading collapse on all tested architectures with \(\ge50\%\) permutable coverage.
\end{enumerate}

\subsection{Limitation: Non-Permutable Architectures}
\label{sec:limitation}

NeuPerm acts only on the permutable layers, so a payload small enough to fit entirely outside those layers evades it: the defense then offers no protection, which is why the analysis above assumes at least some payload rows fall within \(\mathcal{L}_{\mathrm{perm}}\).  That graded case has an extreme form.  An adversary may simply choose a network architecture in which \(\mathcal{L}_{\mathrm{perm}}=\emptyset\) (e.g.\ a purely sequential model with tied weights, or certain RNNs lacking permutation symmetry); recent work further shows that architectures can be constructed to provably lack parameter permutation symmetry while still training normally under standard optimization~\cite{lim2024empirical}.  In that case NeuPerm cannot apply any randomization and \(S=1\).  Practical deployments must therefore ensure models contain at least one large permutable layer to obtain any security benefit.
We further discuss this matter in Section~\ref{sec:adaptive_nonperm}.

\section{Neural Network Steganography Disruption Methods}
\label{sec:nn_stego_disruption}
For comparison, we implement other methods cited by previous works used to mitigate neural network steganography. All the different methods do so by slightly changing the parameters in a way that minimizes model performance degradation. Song et al., and Amit et al. \cite{dl_stego_1,amit_dl_stego_cdr} suggest adding random normal noise to the model parameters - a simple and effective method. Additional options are fine-tuning \cite{amit_dl_stego_cdr,evilmodel1,stegonet}, parameter pruning \cite{maleficnet1}, and quantization \cite{maleficnet1,ran_dl_stego_cdr}.
\subsection{Qualitative Comparison}
See Table \ref{tab:nnsdm} for a comparison of the methods. We consider the following questions: is the method quick, generic, does it not require retraining,  does it not degrade performance, and whether or not it is effective on steganography attacks, StegoNet (SN), EvilModel (EM), and MaleficNet (MN). We mark answers to these questions like so: \emptycircle \; = no, \lefthalfcircle \; = partially, and \fullcircle \; = yes. Adding random noise is the simplest method; it only requires adding two matrices, and hence, it is very quick and the most generic, since it does not depend on different variables like model architecture, parameter type, etc. However, adding too much noise can degrade performance as empirical results show (see Section \ref{sec:exp1}).
Fine-tuning is not quick, since it requires training deep learning models, which might take hours, and it is not generic, since it requires careful manual actions, tailored to the specific model architecture, task, etc.
Parameter pruning and quantization both require fine-tuning to recover lost performance, so they include the downsides of fine-tuning.
MaleficNet \cite{maleficnet1} establishes that adding random noise, parameter pruning, and fine-tuning are not effective methods against it, but they note quantization \textbf{might} work without empirically proving that; our experimental evaluation shows that 8-bit PTQ, which has an acceptable drop in model performance is ineffective against MaleficNet, and that stronger quantization (4-bit, 2-bit) has a drastic impact on model performance (See Section \ref{sec:exp1}, \ref{sec:exp2}).
NeuPerm is quick, as it only requires reordering the axes of the model parameter matrices in place. It does not require fine-tuning, and largely does not degrade model performance thanks to the permutation symmetry property. Finally, NeuPerm is partially generic, since it requires proving the permutation symmetry for different neural network types, and then implementing it as a program; however, there are a relatively small number of neural network architectures (MLP, CNN, RNN, ATTN, etc.), and we cover MLPs, CNNs, and ATTN in this work. We note that NeuPerm's effectiveness against MaleficNet is further conditioned on the target architecture exhibiting high permutable-parameter coverage (in the configurations evaluated here, disruption held when measured coverage was approximately $\geq$50\%; see Section~\ref{sec:limitations}) and on the attacker using a coordinate-dependent extraction procedure as in all currently-deployed neural network steganography attacks (see Section~\ref{sec:adaptive_unshuffle}).

\addtolength{\tabcolsep}{3pt}

\begin{table}[h]
\centering
\caption{Comparison of neural network steganography disruption methods. We report the effectiveness of the methods on prior neural network steganography works: StegoNet (SN), EvilModel (EM), MaleficNet (MN) \cite{stegonet,evilmodel1,maleficnet1}. Effectiveness means whether a method can disrupt neural network steganography to some extent, and is individually indicated by the original authors. Effectiveness against MaleficNet is conditioned on coordinate-dependent extraction and, in the configurations evaluated here, high permutable coverage (approximately $\geq$50\%).}
\label{tab:nnsdm}
\resizebox{\columnwidth}{!}{%
\begin{tabular}{lccccccc}
\toprule
             & Quick & Generic & \begin{tabular}[c]{c}Does not Require\\ Retraining\end{tabular} & \begin{tabular}[c]{c}Does not Degrade\\ Performance\end{tabular} & \multicolumn{3}{c}{Effectiveness} \\ \cmidrule(l){6-8}
Method       &       &         &                                                               &                                                                & SN \cite{stegonet}       & EM \cite{evilmodel1}       & MN \cite{maleficnet1}        \\ \midrule
\rowcolor[HTML]{EFEFEF}
Noise \cite{amit_dl_stego_cdr,ran_dl_stego_cdr}       & \fullcircle     & \fullcircle       & \fullcircle                                                             & \lefthalfcircle                                                              & \fullcircle         & \fullcircle         & \emptycircle         \\
Pruning \cite{maleficnet1}     & \emptycircle     & \emptycircle       & \emptycircle                                                             & \emptycircle                                                              & \fullcircle         & \fullcircle         & \emptycircle         \\
\rowcolor[HTML]{EFEFEF}
Fine-tuning \cite{maleficnet1} & \emptycircle     & \emptycircle       & \emptycircle                                                             & \emptycircle                                                              & \fullcircle         & \fullcircle         & \emptycircle         \\
Quantization \cite{ran_dl_stego_cdr,maleficnet1} & \emptycircle     & \emptycircle       & \emptycircle                                                             & \emptycircle                                                              & \fullcircle         & \fullcircle         & \lefthalfcircle         \\ \midrule
\rowcolor[HTML]{EFEFEF}
NeuPerm (Ours)     & \fullcircle     & \lefthalfcircle      & \fullcircle                                                             & \fullcircle                                                              & \fullcircle         & \fullcircle         & \fullcircle         \\ \bottomrule
\end{tabular}%
}
\end{table}

\addtolength{\tabcolsep}{-3pt}

\subsection{Baseline Disruption Methods Implementation}

We compare NeuPerm against three standard weight-level disruptions: additive Gaussian noise, post-training quantization (PTQ), and random unstructured pruning. All three are applied to a trained model and evaluated with the same pipeline used for NeuPerm.

We do not include fine-tuning as a baseline, since MaleficNet \cite{maleficnet1} already establishes it is ineffective, and it requires task-specific retraining that makes it incomparable to zero-retraining methods.

\subsubsection{Noise Addition}

We inject independent zero-mean Gaussian noise into every trainable parameter of the model. Concretely, we flatten all parameters returned by \texttt{model.parameters()} into a single vector $w \in \mathbb{R}^{N}$, draw $\eta \sim \mathcal{N}(0, \epsilon^{2} I_{N})$, and write $w' = w + \eta$ back into the model via \texttt{torch.\allowbreak nn.\allowbreak utils.\allowbreak vector\_to\_parameters}. The noise scale $\epsilon$ is the only hyperparameter and is swept across several orders of magnitude (typically $10^{-4}$ to $10^{-1}$) to trade off payload disruption against clean-task accuracy. For LLMs, where flattening the full parameter vector is memory-prohibitive, we instead iterate the \texttt{state\_dict} and add noise tensor-by-tensor, which is mathematically equivalent.

\subsubsection{Quantization}

We use weight-only fake quantization (quantize-dequantize, ``QDQ''): weights are quantized to $n$-bit integers and then immediately dequantized back to floating point, so the result is a standard float tensor with reduced effective precision. This lets us reuse the same payload-extraction code as the unquantized baseline.

The procedure has three steps:
\begin{enumerate}
    \item \textbf{BatchNorm fusion.} Because BatchNorm layers absorb a multiplicative scale that would otherwise interact with per-weight quantization error, we first fold them into the preceding Conv/Linear layer. Fusion is performed with PyTorch's FX graph tracer (torch.ao.quantization.quantize\_fx.prepare\_fx) using a no-op \texttt{QConfig}, which triggers the built-in Conv$\to$BN fusion pass without inserting any observers or fake-quant nodes. BN layers that cannot be fused (e.g.\ DenseNet's pre-activation BN$\to$ReLU$\to$Conv pattern) are left untouched and explicitly excluded from quantization.
    \item \textbf{Per-tensor or per-channel scale computation.} For each float weight tensor $w$, we compute the symmetric scale $s = \max(|w|) / q_{\max}$, where $q_{\max} = 2^{n-1} - 1$ and $n$ is the target bit-width. In the per-channel variant, $s$ is computed independently along the output-channel axis (axis 0 for Conv2d/Linear); otherwise a single per-tensor scale is used. Scales are clamped below at $10^{-8}$ to avoid division by zero on all-zero tensors.
    \item \textbf{Quantize-dequantize.} The weight is mapped as
    \[
        \tilde{w} = \mathrm{clip}\!\left(\mathrm{round}(w/s),\; q_{\min},\; q_{\max}\right) \cdot s,
    \]
    with $q_{\min} = -2^{n-1}$, and cast back to its original dtype. Non-float buffers (e.g.\ \texttt{num\_batches\_tracked}) and the BatchNorm parameter keys identified in step~1 are passed through unchanged.
\end{enumerate}

We report results for $n \in \{8, 4, 2\}$ bits, covering both the standard INT8 regime and the more aggressive INT4 and INT2 settings.

\subsubsection{Pruning}

We apply random unstructured pruning to every \texttt{nn.Conv2d} and \texttt{nn.Linear} (this includes all attention and MLP projection layers in transformer architectures) module in the network, using

\noindent torch.nn.utils.prune.random\_unstructured with a target sparsity \texttt{amount} $\in [0,1]$. This zeros out a uniformly-random fraction of the weight entries of each layer independently, leaving biases and normalization parameters untouched. The pruning is unstructured (individual scalar weights, not whole channels) and random (not magnitude-based), which matches the adversarial assumption that the attacker does not know which weights carry the hidden payload. As with noise, we sweep the sparsity \texttt{amount} to trace out the disruption/accuracy trade-off curve.

\section{Datasets}
\label{sec:dataset}
To evaluate NeuPerm, we aim to analyze two key properties:
(1) Model performance after applying NeuPerm.
(2) Survivability of neural network steganography (EvilModel, MaleficNet) after applying NeuPerm.

\noindent \textbf{Pre-trained CNN models:} We used 5 well-known image classification CNN architectures: DenseNet, EfficientNet, MobileNet,
ResNet, and VGG. In particular, we use the DenseNet121, EfficientNet-B0, EfficientNet-B4, MobileNet-V2, MobileNet-V3 (Small), ResNet50, ResNet101, VGG11, and VGG16 variants.
For evaluating model performance, we use pre-trained PyTorch models trained on the ImageNet12 dataset~\cite{russakovsky2015imagenet}, and we evaluate them on the validation split.

\noindent \textbf{Pre-trained LLM models:} We use the meta-llama/Llama-3.2-1B-Instruct and Qwen/Qwen2.5-1.5B-Instruct HuggingFace models with bfloat16 parameters; the sole exception is the post-training-quantization experiment, which loaded both models in float16. We evaluate them on three complementary benchmarks that probe the model at increasing distance from a task decision: SQuAD~v1.1~\cite{rajpurkar-etal-2016-squad} (extractive span selection, scored by token-level F1), BoolQ~\cite{clark2019boolq} (binary yes/no classification, scored by accuracy), and WikiText-2~\cite{merity2016pointer} (raw language-model perplexity, lower is better); the last reads the full next-token distribution directly, with no chat template or generation, making it the most direct probe of the input-output function NeuPerm preserves. For SQuAD and BoolQ we evaluate 200 validation examples per run using each model's native chat template with greedy decoding (SQuAD: a shortest-answer-span instruction, up to 32~new tokens; BoolQ: a yes/no instruction, up to 4~new tokens, classifying the first continuation token as ``yes'', ``no'', or - counted wrong, neither). For WikiText-2 we score the raw next-token likelihood over the full test split with a 1024-token sliding window (stride 512), accumulating negative log-likelihoods in float32 and reporting the exponential of their mean.

\noindent \textbf{Reproduced MaleficNet stegomodels:} We reproduce MaleficNet \cite{maleficnet1,maleficnet2} using the authors' released code, embedding real malware payloads into model parameters with their error-correcting, spread-spectrum scheme. We apply this attack to two groups of models. The first \emph{reproduces} the models from the original MaleficNet papers: the DenseNet121, ResNet50/101, and VGG11/16 CNNs on the ImageNet~\cite{russakovsky2015imagenet} and CIFAR-10~\cite{krizhevsky2009cifar} datasets of \cite{maleficnet1}, together with the Llama-3.2-1B LLM introduced in \cite{maleficnet2}. The second \emph{extends} the same attack to architectures the original papers did not consider, to test NeuPerm's generality: the modern EfficientNet-B0/B4 and MobileNet-V2/V3~(Small) CNNs, and the Qwen2.5-1.5B LLM.
For the malware, we use 9 samples from TheZoo \cite{thezoo}: Stuxnet, Destover, Asprox, Bladabindi, Zeus-Bank, EquationDrug, Kovter, Cerber, and Ardamax. Not every payload fits in every model, since MaleficNet's spread-spectrum capacity scales with parameter count.
In total this yields \textbf{78} CNN MaleficNet stegomodels (across ImageNet and CIFAR-10) and \textbf{18} LLM MaleficNet stegomodels (Llama-3.2-1B and Qwen2.5-1.5B, each embedded with the 9 payloads).

\section{Experimental Results}
\label{sec:experiments}
In this section, we present experimental results of NeuPerm and related baseline methods (see Section \ref{sec:nn_stego_disruption}). We evaluate the methods in terms of 2 aspects: \textbf{(1) effect on model performance:} we consider methods with the least difference in model performance before and after applying them, the most optimal; we evaluate the models on validation tasks for this purpose. \textbf{(2) effectiveness against neural network steganography:} MaleficNet \cite{maleficnet1,maleficnet2} is our primary target and the harder problem: unlike earlier attacks it embeds the payload with error-correcting resilience, and no prior defense is proven to work against it (its authors only conjecture without proof that model quantization might). We therefore center our disruption evaluation on MaleficNet. The remaining non-error-correcting attacks such as Song et al. \cite{dl_stego_1}, StegoNet \cite{stegonet}, and EvilModel \cite{evilmodel1,evilmodel2} rely on the parameters staying byte-exact and are broken by any parameter change, as argued in Section \ref{sec:fsa_noecc}; we confirm this empirically on the EvilModel attack, thereby demonstrating generalization beyond MaleficNet. See Section \ref{sec:dataset} for the data we used for our experiments (models, benchmarks, etc.). Experiment 1 (Section \ref{sec:exp1}) evaluates model performance, Experiment 2 (Section \ref{sec:exp2}) evaluates effectiveness against MaleficNet, and Experiment 3 (Section \ref{sec:exp3}) evaluates NeuPerm on the EvilModel attack.

Throughout this section, wherever a table cell is written as a mean $\pm$ standard deviation, the value is the mean over independent seeded replicates, and the reported spread reflects the random draw of the perturbation or the permutation rather than evaluation nondeterminism, that is, the run-to-run variation of evaluating one fixed set of weights, which we measured at exactly zero (Appendix~\ref{sec:repro}). The number of replicates $n$ varies by table and is stated in each caption. The seeding scheme, the definition of one replicate, and the 95\% confidence-interval rule are given in Appendix~\ref{sec:repro}.



\subsection{Experiment 1: NeuPerm Does Not Degrade Performance}
\label{sec:exp1}
In this experiment, we analyze the effects of NeuPerm on model performance. Even though we proved NeuPerm does not change the neural network computation theoretically (see Section \ref{sec:neuperm_proof}), in practice, GPU computations are not fully deterministic, and floating numbers can not perfectly represent all real numbers; hence, we also assert that \textbf{NeuPerm does not degrade model performance} empirically.

We note that the accuracy and SNR experiments were conducted on GPU using PyTorch's default (non-deterministic) CUDA kernels; the host used for each results table is given in the measurement-environment table of Section~\ref{sec:repro}. The minor score fluctuations observed with NeuPerm are attributable to non-deterministic floating-point reduction order in GPU operations, and may vary slightly across hardware platforms or when deterministic mode is enabled. The effect is negligible in all cases we tested.

We analyze CNNs and two LLMs.
See Table \ref{tab:exp1_acc}. Our reproduced CNN models achieve similar accuracy on the ImageNet12 validation set. We can see NeuPerm (full) causes no difference or very negligible changes in model accuracy for all the chosen CNN architectures. The nine CNN architectures all got up to a 0.0048~percentage-point (pp) NeuPerm-minus-baseline accuracy fluctuation, which is negligible. Compared with NeuPerm, adding random normal noise to the parameters with a standard deviation of $0.0001$ does degrade performance slightly, and as the standard deviation increases, the model performance is completely eroded. Pruning even 1\% of the model parameters without retraining to regain accuracy has a dramatic negative effect. Finally, 8-bit quantization is shown to cause a fluctuation of up to 1.5\% accuracy, in all models but the MobileNet-V3 (Small) model, which is acceptable, but stronger quantization (4-bit, 2-bit) is too disruptive and destroys model performance, making these methods unviable.
Across the stochastic baselines, random unstructured pruning and mid-range noise place the model on a performance cliff, so the outcome swings with which weights the draw happens to hit, reaching a standard deviation of 22.204~pp in the worst case (EfficientNet-B0, pruning at 1\%). NeuPerm's own standard deviation stays at or below 0.005~pp on every architecture, so it is reproducible to within floating-point reduction order where the baselines it is compared against are unreliable from run to run.
We can conclude that NeuPerm does not affect CNN model performance as the theory suggests.

\begin{table*}[!t]
\centering
\caption{Mean validation accuracy (\%) of select PyTorch pre-trained CNN models on the ImageNet12 dataset: the accuracy cited on PyTorch's site, the accuracy reproduced on our machine, and the accuracy after each disruption technique, NeuPerm (ours), additive random normal noise, parameter pruning, and quantization. Stochastic conditions (NeuPerm, all four noise levels, both pruning ratios) are reported as mean $\pm$ standard deviation over $n=5$ independent seeded replicates; deterministic conditions (the reproduced baseline and the three quantization settings) are evaluated once ($n=1$) and carry no $\pm$ term. The PyTorch Reported Accuracy column is quoted, not measured by us. \textbf{Only full NeuPerm, adding random noise with a 0.0001 standard deviation, and 8-bit quantization have acceptable performance drops.}}
\label{tab:exp1_acc}
\resizebox{\textwidth}{!}{%
\begin{tabular}{|l|cc|c|cccc|cc|ccc|}
\hline
                & \begin{tabular}[c]{@{}c@{}}PyTorch\\ Reported\\ Accuracy\end{tabular} & \begin{tabular}[c]{@{}c@{}}Our\\ Reproduced\\ Accuracy\end{tabular} & NeuPerm          & \multicolumn{4}{c|}{Random Normal Noise}                                                              & \multicolumn{2}{c|}{Prune}                  & \multicolumn{3}{c|}{Quantization}                    \\
                &                                                                       &                                                                     &                  & \multicolumn{4}{c|}{Standard Deviation}                                                               & \multicolumn{2}{c|}{Pruning Ratio}          & \multicolumn{3}{c|}{Bit Width}                       \\
                &                                                                       &                                                                     &                  & 0.0001           & 0.001            & 0.01            & 0.1             & 1\%              & 5\%              & 8                & 4                & 2             \\ \hline
DenseNet121     & \textbf{74.434} & \textbf{74.438} & \textbf{74.434 $\pm$ 0.003} & \textbf{74.405 $\pm$ 0.019} & \textbf{74.146 $\pm$ 0.087} & 8.950 $\pm$ 4.455 & 0.100 $\pm$ 0.000 & 69.960 $\pm$ 0.754 & 37.610 $\pm$ 3.368 & \textbf{74.342} & 19.876 & 0.100 \\
\rowcolor[HTML]{EFEFEF}
ResNet50        & \textbf{80.858} & \textbf{80.846} & \textbf{80.843 $\pm$ 0.003} & \textbf{80.824 $\pm$ 0.013} & 64.325 $\pm$ 18.176 & 0.119 $\pm$ 0.076 & 0.101 $\pm$ 0.005 & 60.361 $\pm$ 5.908 & 5.955 $\pm$ 9.012 & \textbf{80.718} & 0.690 & 0.100 \\
ResNet101       & \textbf{81.886} & \textbf{81.900} & \textbf{81.902 $\pm$ 0.001} & \textbf{81.886 $\pm$ 0.030} & 78.694 $\pm$ 3.754 & 0.104 $\pm$ 0.017 & 0.100 $\pm$ 0.000 & 74.996 $\pm$ 2.422 & 8.940 $\pm$ 4.079 & \textbf{81.748} & 10.364 & 0.100 \\
\rowcolor[HTML]{EFEFEF}
VGG11           & \textbf{69.020} & \textbf{69.048} & \textbf{69.046 $\pm$ 0.003} & \textbf{69.034 $\pm$ 0.016} & \textbf{68.950 $\pm$ 0.020} & 60.801 $\pm$ 0.381 & 0.098 $\pm$ 0.015 & 65.317 $\pm$ 0.727 & 39.795 $\pm$ 5.606 & \textbf{68.998} & 59.372 & 0.100 \\
VGG16           & \textbf{71.592} & \textbf{71.586} & \textbf{71.581 $\pm$ 0.002} & \textbf{71.578 $\pm$ 0.010} & \textbf{71.527 $\pm$ 0.040} & 62.124 $\pm$ 1.000 & 0.098 $\pm$ 0.013 & 69.678 $\pm$ 2.144 & 60.892 $\pm$ 3.757 & \textbf{71.634} & 67.860 & 0.100 \\
\rowcolor[HTML]{EFEFEF}
EfficientNet-B0 & \textbf{77.692} & \textbf{77.692} & \textbf{77.692 $\pm$ 0.001} & \textbf{77.689 $\pm$ 0.014} & \textbf{77.611 $\pm$ 0.049} & 70.711 $\pm$ 0.659 & 0.094 $\pm$ 0.015 & 17.136 $\pm$ 22.204 & 0.175 $\pm$ 0.067 & \textbf{76.914} & 0.110 & 0.100 \\
EfficientNet-B4 & \textbf{83.384} & \textbf{83.392} & \textbf{83.396 $\pm$ 0.002} & \textbf{83.396 $\pm$ 0.005} & \textbf{83.396 $\pm$ 0.012} & \textbf{83.191 $\pm$ 0.189} & 0.093 $\pm$ 0.006 & 74.596 $\pm$ 5.070 & 3.926 $\pm$ 1.827 & \textbf{83.170} & 0.490 & 0.122 \\
\rowcolor[HTML]{EFEFEF}
MobileNetV2     & \textbf{72.154} & \textbf{72.148} & \textbf{72.145 $\pm$ 0.005} & \textbf{72.134 $\pm$ 0.021} & \textbf{71.964 $\pm$ 0.084} & 41.254 $\pm$ 4.050 & 0.103 $\pm$ 0.012 & 11.882 $\pm$ 6.824 & 0.124 $\pm$ 0.024 & \textbf{71.186} & 0.120 & 0.112 \\
MobileNetV3 (S) & \textbf{67.668} & \textbf{67.672} & \textbf{67.669 $\pm$ 0.005} & \textbf{67.649 $\pm$ 0.013} & \textbf{67.478 $\pm$ 0.063} & 55.531 $\pm$ 2.686 & 0.110 $\pm$ 0.009 & 7.558 $\pm$ 6.858 & 0.111 $\pm$ 0.025 & 65.844 & 0.094 & 0.100 \\
\hline
\end{tabular}%
}

\end{table*}

Moving on to the LLMs (Llama-3.2-1B-Instruct and Qwen2.5-1.5B-Instruct, bfloat16). See Table \ref{tab:exp1_llm}. NeuPerm caused a 0.3$\pm$0.5 F1 score difference on Llama-3.2-1B-Instruct and a $-$0.03$\pm$0.1 F1 score difference on Qwen2.5-1.5B-Instruct relative to their respective baselines on SQuAD. To verify this generalizes beyond extractive QA, we additionally evaluate both LLMs on BoolQ (yes/no question answering), reported in the middle block of Table~\ref{tab:exp1_llm}. NeuPerm again preserves accuracy on both models to within seed-level noise (the NeuPerm row sits within $\pm$1\% of the base score on both LLMs), confirming that the small fluctuations we observe are non-deterministic floating-point reduction effects rather than a degradation specific to one task. As the most direct test of function preservation, we further evaluate both LLMs on WikiText-2 perplexity (raw next-token likelihood, with no chat template and no generation), reported in the bottom block of Table~\ref{tab:exp1_llm}. Here NeuPerm changes perplexity by only $+0.0005$ on Llama-3.2-1B-Instruct and $+0.0004$ on Qwen2.5-1.5B-Instruct - below the $\sigma{=}10^{-4}$ noise floor and consistent with float-reduction non-determinism rather than any functional change, exactly as the permutation-equivalence theory predicts. As we anticipated, the theory guarantees that the neural network computes the same function, but it relies on various factors like the associativity of addition, etc., which are not modeled perfectly in computing, especially with GPU multiprocessing, and the fact that computer float values can not represent all mathematically possible real numbers. Nevertheless, we argue this is a negligible change in performance, and adding random noise or parameter pruning produced results with substantial variability across all three benchmarks, limiting the reliability of these methods compared with NeuPerm, which had a small standard deviation. Similarly to the CNNs, 8-bit quantization caused acceptable performance fluctuations, and stronger quantization (4-bit, 2-bit) destroyed the model's performance on SQuAD and BoolQ and inflated WikiText-2 perplexity by orders of magnitude.

\addtolength{\tabcolsep}{3pt}

\begin{table*}[!t]
\centering
\caption{Mean performance $\pm$ standard deviation of the Llama-3.2-1B-Instruct and Qwen2.5-1.5B-Instruct LLMs (bfloat16) under steganography-disruption techniques on three complementary benchmarks: SQuAD (F1-Score \%, top block), BoolQ (accuracy \%, middle block), and WikiText-2 (perplexity, bottom block). We report each base model evaluated on our machine, then the same model after applying NeuPerm (ours), adding random normal noise, parameter pruning, and quantization. Every condition, including the deterministic quantization conditions, is repeated over $n=5$ independent seeded replicates, and those quantization conditions showed zero standard deviation across the five seeds. A WikiText-2 entry of $\infty$ marks a configuration whose perplexity diverged (numerically overflowed). \textbf{For SQuAD and BoolQ higher is better, and for WikiText-2 perplexity lower is better. Only Full NeuPerm, adding random noise with 0.0001 standard deviation, and 8-bit quantization have acceptable performance fluctuations; NeuPerm had the scores closest to the baseline on both LLMs and all three benchmarks.}}
\label{tab:exp1_llm}
\resizebox{\textwidth}{!}{%
\begin{tabular}{lcccccccccll}
\hline
Model & Base                 & NeuPerm              & \multicolumn{4}{c}{Random Normal Noise}        & \multicolumn{2}{c}{Prune}         & \multicolumn{3}{c}{Quantization}                                                         \\
      & \multicolumn{1}{l}{} & \multicolumn{1}{l}{} & \multicolumn{4}{c}{Standard Deviation}         & \multicolumn{2}{c}{Pruning Ratio} & \multicolumn{3}{c}{Bit Width}                                                            \\ \cline{4-12}
      &                      &                      & 0.0001               & 0.001      & 0.01 & 0.1 & 1\%             & 5\%             & 8                                        & \multicolumn{1}{c}{4} & \multicolumn{1}{c}{2} \\ \hline
\multicolumn{12}{l}{\textbf{SQuAD (F1-Score \%)}} \\ \hline
Llama-3.2-1B-Instruct  & \textbf{77.05}       & \textbf{77.35 ± 0.5} & \textbf{77.59 ± 0.6} & 76.34 ± 2.6 & 0    & 0   & 73.91 ± 2.9     & 46.07 ± 4.7     & \multicolumn{1}{l}{\textbf{77.05 ± 0.0}} & 27.62 ± 0.0           & 0                     \\
Qwen2.5-1.5B-Instruct  & \textbf{92.12}       & \textbf{92.09 ± 0.1} & \textbf{92.16 ± 0.2} & 91.37 ± 0.7 & 0.21 & 0   & 88.85 ± 2.8     & 49.65 ± 12.4    & \multicolumn{1}{l}{\textbf{91.00 ± 0.0}} & 0.38 ± 0.0            & 0                     \\ \hline
\multicolumn{12}{l}{\textbf{BoolQ (Accuracy \%)}} \\ \hline
Llama-3.2-1B-Instruct  & \textbf{56.50}       & \textbf{56.70 ± 0.9} & \textbf{56.50 ± 1.2} & 60.10 ± 6.2 & 0           & 0.40 ± 0.7  & 60.60 ± 5.6          & 47.40 ± 17.1    & \multicolumn{1}{l}{\textbf{54.50 ± 0.0}} & 36.50 ± 0.0           & 0                     \\
Qwen2.5-1.5B-Instruct  & \textbf{74.50}       & \textbf{74.60 ± 0.6} & \textbf{74.70 ± 0.8} & \textbf{74.80 ± 1.4} & 0.60 ± 0.7  & 0.10 ± 0.2  & \textbf{73.80 ± 2.1} & 65.60 ± 10.2    & \multicolumn{1}{l}{\textbf{75.00 ± 0.0}} & 0                     & 0                     \\ \hline
\multicolumn{12}{l}{\textbf{WikiText-2 (Perplexity, lower is better)}} \\ \hline
Llama-3.2-1B-Instruct  & \textbf{12.368}      & \textbf{12.369 ± 0.001} & \textbf{12.372 ± 0.005} & 13.042 ± 0.067 & $\infty$ & $\infty$ & 14.250 ± 1.028 & 29.693 ± 2.455 & \multicolumn{1}{l}{\textbf{12.404 ± 0.0}} & 49.514 ± 0.0          & $\infty$   \\
Qwen2.5-1.5B-Instruct  & \textbf{9.159}       & \textbf{9.159 ± 0.001}  & \textbf{9.161 ± 0.004}  & 9.384 ± 0.027  & $\infty$ & $\infty$ & 10.500 ± 0.358 & 23.351 ± 1.970 & \multicolumn{1}{l}{\textbf{9.202 ± 0.0}}  & $\infty$   & $\infty$   \\ \hline
\end{tabular}%
}

\end{table*}

\addtolength{\tabcolsep}{-3pt}

\subsection{Experiment 2: NeuPerm Disrupts MaleficNet Extraction on the Evaluated High-Coverage Architectures}
\label{sec:exp2}
In this experiment, we analyze how effectively NeuPerm disrupts CNN and LLM MaleficNet stegomodels, and we compare it to adding random noise, and quantization, which are other comparable disruption techniques (see Section \ref{sec:nn_stego_disruption}). We chose to compare only against methods which caused acceptable performance fluctuations in Experiment 1, as they are the methods which are directly comparable to NeuPerm, so we chose adding random noise with 0.0001 standard deviation, and 8-bit quantization.
According to the MaleficNet paper \cite{maleficnet2}, given a MaleficNet stegomodel, it is proven that if the signal-to-noise ratio (SNR) is lower than 1, then the payload can not be successfully extracted. Therefore, we use SNR to measure how effective applied techniques are, and we regard SNR values of less than 1 to mean the attack has been successfully disrupted and the payload can not be extracted.
We compute the SNR using the extraction and evaluation code provided by the MaleficNet authors \cite{maleficnet2}.
See Table \ref{tab:exp2_in12}, we focus on the CNN ImageNet12 and LLM MaleficNet stegomodels (see Section \ref{sec:dataset}). We can see that full NeuPerm caused the SNR to become negative in every tested architecture, showing by the SNR${<}1$ criterion that the attack was successfully disrupted and the payload can not be extracted.

The comparable methods, such as adding random noise with a standard deviation of 0.0001, and 8-bit quantization had little to no effect on the SNR. This aligns with the empirical results shown in MaleficNet \cite{maleficnet1,maleficnet2}, which additionally asserts that fine-tuning and parameter pruning are also ineffective; additionally showing that 8-bit quantization is also insufficient.

\begin{figure*}[!t]
    \centering
    \includegraphics[width=0.8\linewidth]{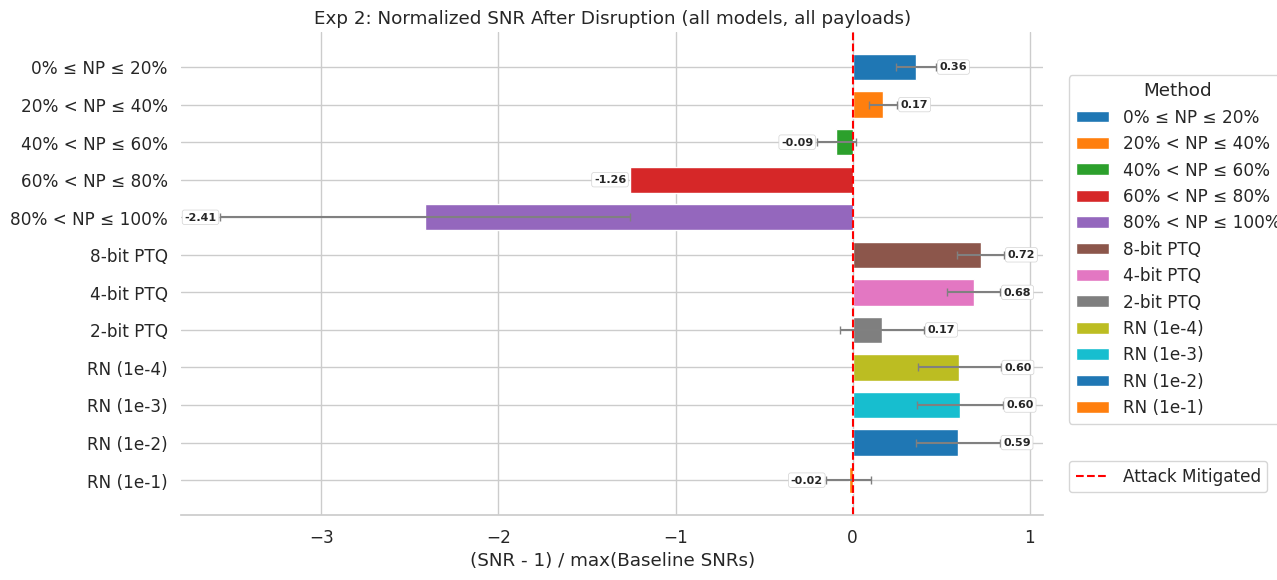}
        \caption{Effectiveness of disruption techniques on CNN MaleficNet stegomodels (ImageNet12 and CIFAR-10 combined). The $y$-axis shows the mean normalized SNR difference: for each stegomodel, we compute $(\text{SNR} - 1) / \max(\text{Baseline SNRs})$, the gap to the SNR${=}1$ extraction threshold normalized by the largest baseline SNR, then average across all architecture-malware combinations. \textbf{Values below 0 indicate SNR ${<}1$, below the extraction threshold, i.e., the attack was disrupted.} The $x$-axis groups the disruption methods. NeuPerm (NP) is tested at varying coverage levels: ``$a\% \le \text{NP} \le b\%$'' denotes NeuPerm strategies that permuted between $a\%$ and $b\%$ of the model's parameters (see Table~\ref{tab:neuperm_cnns} for per-architecture coverage). Error bars denote $\pm 1$ sample standard deviation across the measurements contributing to each point, where each measurement is one stegomodel under one permutation configuration. Random Noise (RN) is tested at increasing standard deviations $f \in \{0.0001, 0.001, 0.01, 0.1\}$. 8/4/2-bit Post-Training Quantization (X-bit PTQ) are also shown. All methods are applied to the same stegomodel set (Section~\ref{sec:dataset}). \textbf{Takeaway:} NeuPerm's disruption strengthens monotonically with permutable coverage: high-coverage configurations drive the mean normalized SNR difference well below $0$, while low-coverage ($\le 40\%$) NeuPerm do not. Random Noise succeeds only at $f=0.1$, and then only marginally (mean $-0.023$), a setting that destroys model accuracy (see Section~\ref{sec:exp1}). PTQ has no meaningful effect.}
    \label{fig:exp2_diff_new}
\end{figure*}

We also provide a more in-depth comparison of NeuPerm (NP) with varying degrees of percent of parameters changed, adding random noise with increasing standard deviation, and 8/4/2-bit PTQ. See Figure \ref{fig:exp2_diff_new}. As NeuPerm coverage increases, payload SNR decreases, as expected. In the MaleficNet configurations evaluated in this study, payload SNR fell below the extraction threshold when measured permutable-parameter coverage was approximately 50\% or higher. We adopt this figure as an empirical rule of thumb for the evaluated models. What ultimately determines disruption is whether the payload-bearing coordinates are permuted, and higher permutable coverage makes that increasingly likely, which is why coverage predicts disruption so reliably on these architectures. The exact figure is model-specific because it also depends on the payload, payload size, layer-selection strategy, and error-correcting parameters. In our tests, 8/4/2-bit PTQ did not disrupt MaleficNet extraction at any setting. Adding random noise disrupts extraction only at a standard deviation of at least 0.1, and we recall that random noise beyond a standard deviation of 0.0001 significantly hinders model performance. Random noise and PTQ are therefore not practical methods for disrupting MaleficNet extraction, whereas on the evaluated architectures NeuPerm is, at the permutable-parameter coverage those models admit.
\addtolength{\tabcolsep}{2pt}

\begin{table*}[!t]
\centering
\caption{SNR of ImageNet12 CNN and LLM (Llama-3.2-1B, Qwen2.5-1.5B) MaleficNet models before and after applying disruption techniques. Baseline (B) is the SNR of the model after attacking it with MaleficNet. Random Noise (RN) is the SNR after adding random normal noise (baseline method \cite{amit_dl_stego_cdr}) with a standard deviation of 0.0001. 8-bit Quantization (Q8) is the SNR after applying 8-bit quantization to the parameters (baseline method \cite{maleficnet1,ran_dl_stego_cdr}). NeuPerm (NP) (our suggested method) is the SNR after applying full NeuPerm. \textbf{Lower SNR is better, SNR less than 1 means the attack is disrupted}. A "-" marks a model-payload combination we did not measure, whether because the payload exceeds that model's spread-spectrum capacity (Section~\ref{sec:dataset}) or because the cell was not run. Each NeuPerm cell is the mean $\pm$ standard deviation over 10 independent permutation draws. All columns round to one decimal place. Measurement provenance, the payload basis of each cell, agreement tolerances, and reconciliation with previously published single draws are given in Appendix~\ref{sec:repro}. \textbf{Takeaway:} On the evaluated architectures NeuPerm disrupts MaleficNet extraction, as designed, at the high permutable coverage those models admit (approximately $\geq$50\% in these configurations), whereas quantization and additive random noise do not.}
\label{tab:exp2_in12}
\resizebox{\textwidth}{!}{%
\begin{tabular}{lcccccccccccccccccccc}
\hline
           & \multicolumn{4}{c}{MobileNet-V2} & \multicolumn{4}{c}{MobileNet-V3 (S)} & \multicolumn{4}{c}{EfficientNet-B0} & \multicolumn{4}{c}{DenseNet121}                          & \multicolumn{4}{c}{EfficientNet-B4} \\ \cline{2-21} 
Malware    & B      & RN     & Q8     & NP    & B     & RN   & Q8   & NP             & B       & RN      & Q8     & NP     & B   & RN  & Q8                          & NP             & B       & RN      & Q8     & NP     \\ \hline
Stuxnet    & 4.9 & 4.9 & 4.9 & \textbf{-6.8 $\pm$ 0.8} & 3.8 & 3.8 & 3.8 & \textbf{-3.3 $\pm$ 0.9} & 5.9 & 5.9 & 5.9 & \textbf{-9.8 $\pm$ 1.1} & 5.1 & 5.1 & 5.1 & \textbf{-9.0 $\pm$ 1.0} & 5.1 & 5.1 & 5.1 & \textbf{-18.5 $\pm$ 2.0} \\
\rowcolor[HTML]{EFEFEF} 
Destover   & - & - & - & - & - & - & - & - & - & - & - & - & 4.2 & 4.2 & \cellcolor[HTML]{EFEFEF}4.2 & \textbf{-11.5 $\pm$ 1.6} & 4.4 & 4.4 & 4.4 & \textbf{-22.1 $\pm$ 5.0} \\
Asprox     & - & - & - & - & - & - & - & - & - & - & - & - & - & - & - & - & 4.7 & 4.7 & 4.7 & \textbf{-20.2 $\pm$ 4.1} \\
\rowcolor[HTML]{EFEFEF} 
Bladabindi & - & - & - & - & - & - & - & - & - & - & - & - & - & - & - & - & 4.8 & 4.8 & 4.8 & \textbf{-22.1 $\pm$ 4.8} \\
\end{tabular}%

}
\par\vspace{3pt}
\resizebox{\textwidth}{!}{

\begin{tabular}{lcccccccccccccccccccccccc}
\hline
           & \multicolumn{4}{c}{ResNet50}                             & \multicolumn{4}{c}{ResNet101}                            & \multicolumn{4}{c}{VGG11}                                & \multicolumn{4}{c}{VGG16} & \multicolumn{4}{c}{Llama-3.2-1B} & \multicolumn{4}{c}{Qwen2.5-1.5B}                       \\ \cline{2-25}
Malware    & B   & RN  & Q8                          & NP             & B   & RN  & Q8                          & NP             & B   & RN  & Q8                          & NP             & B  & RN  & Q8 & NP        & B   & RN  & Q8                          & NP            & B   & RN  & Q8                          & NP            \\ \hline
Stuxnet    & 1.9 & 1.9 & 1.9 & \textbf{-23.9 $\pm$ 4.9} & 5.4 & 5.4 & 5.4 & \textbf{-14.8 $\pm$ 2.3} & 7.7 & 7.7 & 7.7 & \textbf{-30.3 $\pm$ 7.0} & 7.7 & 7.7 & 7.7 & \textbf{-27.3 $\pm$ 5.6} & 7.5 & 7.5 & 7.5 & \textbf{-1.8 $\pm$ 0.5} & 6.6 & 6.6 & 6.6 & \textbf{-2.4 $\pm$ 0.5} \\
\rowcolor[HTML]{EFEFEF} 
Destover   & 1.9 & 1.9 & \cellcolor[HTML]{EFEFEF}1.9 & \textbf{-23.9 $\pm$ 5.9} & 5.1 & 5.1 & \cellcolor[HTML]{EFEFEF}5.1 & \textbf{-16.5 $\pm$ 2.2} & 8.1 & 8.1 & \cellcolor[HTML]{EFEFEF}8.1 & \textbf{-30.6 $\pm$ 11.6} & 7.4 & 7.4 & 7.4 & \textbf{-29.9 $\pm$ 13.2} & 7.7 & 7.7 & \cellcolor[HTML]{EFEFEF}7.7 & \textbf{-2.0 $\pm$ 0.5} & 6.6 & 6.6 & \cellcolor[HTML]{EFEFEF}6.6 & \textbf{-2.7 $\pm$ 0.5} \\
Asprox     & 1.3 & 1.3 & 1.3 & \textbf{-22.8 $\pm$ 4.6} & 5.0 & 5.0 & 5.0 & \textbf{-15.5 $\pm$ 3.0} & 7.2 & 7.2 & 7.2 & \textbf{-26.6 $\pm$ 11.0} & 7.3 & 7.3 & 7.3 & \textbf{-28.7 $\pm$ 8.2} & 6.9 & 6.9 & 6.9 & \textbf{-1.9 $\pm$ 0.4} & 6.8 & 6.8 & 6.8 & \textbf{-3.0 $\pm$ 0.5} \\
\rowcolor[HTML]{EFEFEF} 
Bladabindi & 2.2 & 2.2 & \cellcolor[HTML]{EFEFEF}2.2 & \textbf{-25.9 $\pm$ 11.7} & 4.9 & 4.9 & \cellcolor[HTML]{EFEFEF}4.9 & \textbf{-15.7 $\pm$ 2.8} & 7.7 & 7.7 & \cellcolor[HTML]{EFEFEF}7.7 & \textbf{-21.7 $\pm$ 6.9} & 7.5 & 7.5 & 7.5 & \textbf{-25.4 $\pm$ 6.5} & 7.8 & 7.8 & \cellcolor[HTML]{EFEFEF}7.8 & \textbf{-1.3 $\pm$ 0.4} & 6.7 & 6.7 & \cellcolor[HTML]{EFEFEF}6.7 & \textbf{-2.2 $\pm$ 0.6} \\
Zeus-Bank  & - & - & - & - & 4.8 & 4.8 & 4.8 & \textbf{-16.9 $\pm$ 4.5} & 7.4 & 7.4 & 7.5 & \textbf{-25.7 $\pm$ 7.5} & 8.0 & 8.0 & 8.0 & \textbf{-25.5 $\pm$ 8.5} & 7.6 & 7.6 & 7.6 & \textbf{-1.5 $\pm$ 0.4} & 6.6 & 6.6 & 6.6 & \textbf{-2.1 $\pm$ 0.6} \\
\rowcolor[HTML]{EFEFEF} 
Eq.Drug    & - & - & - & - & 3.9 & 3.9 & \cellcolor[HTML]{EFEFEF}3.9 & \textbf{-16.3 $\pm$ 4.5} & 7.1 & 7.1 & \cellcolor[HTML]{EFEFEF}7.1 & \textbf{-25.4 $\pm$ 8.9} & 6.7 & 6.7 & 6.7 & \textbf{-25.3 $\pm$ 6.4} & 7.0 & 7.0 & \cellcolor[HTML]{EFEFEF}7.0 & \textbf{-1.7 $\pm$ 0.4} & 6.7 & 6.7 & \cellcolor[HTML]{EFEFEF}6.7 & \textbf{-1.9 $\pm$ 0.5} \\
Kovter     & - & - & - & - & 4.2 & 4.2 & 4.2 & \textbf{-20.0 $\pm$ 5.4} & 7.1 & 7.1 & 7.1 & \textbf{-28.5 $\pm$ 6.9} & 7.3 & 7.3 & 7.3 & \textbf{-30.3 $\pm$ 8.7} & 7.8 & 7.8 & 7.8 & \textbf{-1.7 $\pm$ 0.5} & 7.5 & 7.5 & 7.5 & \textbf{-2.6 $\pm$ 0.7} \\
\rowcolor[HTML]{EFEFEF} 
Cerber     & - & - & - & - & 4.4 & 4.4 & \cellcolor[HTML]{EFEFEF}4.4 & \textbf{-21.6 $\pm$ 8.7} & 6.9 & 6.9 & \cellcolor[HTML]{EFEFEF}6.9 & \textbf{-30.2 $\pm$ 8.5} & 7.1 & 7.1 & 7.1 & \textbf{-31.3 $\pm$ 13.2} & 7.0 & 7.0 & \cellcolor[HTML]{EFEFEF}7.0 & \textbf{-2.2 $\pm$ 0.5} & 7.0 & 7.0 & \cellcolor[HTML]{EFEFEF}7.0 & \textbf{-3.1 $\pm$ 0.6} \\
Ardamax    & - & - & - & - & - & - & - & - & 5.9 & 5.9 & 5.9 & \textbf{-32.2 $\pm$ 9.5} & 6.8 & 6.8 & 6.8 & \textbf{-27.5 $\pm$ 6.6} & 7.2 & 7.2 & 7.2 & \textbf{-1.5 $\pm$ 0.7} & 6.7 & 6.7 & 6.7 & \textbf{-3.6 $\pm$ 0.6} \\
\hline
\end{tabular}%

}

\end{table*}

\addtolength{\tabcolsep}{-2pt}

\subsection{Experiment 3: Disrupting a Byte-Exact Attack (EvilModel)}
\label{sec:exp3}

Experiment~2 targets the hardest case: MaleficNet's error-correcting, spread-spectrum payload. To confirm that NeuPerm generalizes beyond MaleficNet, we additionally evaluate NeuPerm on EvilModel~2.0~\cite{evilmodel1,evilmodel2}, a high-capacity byte-substitution attack with no error-correction: we embed a benign random byte payload of known content into the model parameters, then attempt to extract and verify it both before and after applying NeuPerm.

Since the EvilModel authors did not release their code, we reimplement EvilModel's byte-substitution encoding on \texttt{float32} weights (author reimplementation) and evaluate its \emph{half substitution} scheme ($X{=}16$, preserving the leading two bytes of each parameter and overwriting the low two), the scheme EvilModel reports as its strongest; the other published schemes are out of scope for reasons documented in Appendix~\ref{sec:evilmodel_reimpl}. For each of the eight evaluated CNNs (VGG11, VGG16, ResNet50, ResNet101, DenseNet121, EfficientNet-B0, EfficientNet-B4, and MobileNetV2), we embed a benign random byte payload of known content filling $25\%$ of the permutable-site weight capacity, and store the payload length and its SHA-256 digest in the carrier bias as an integrity header, following EvilModel's integrity convention. We extract and verify the payload on the stego-model, apply NeuPerm (5 seeded repeats), and re-extract from the same fixed coordinates; success is the byte-exact SHA-256 check, and we also report the bit-error rate (BER). The reimplementation's faithfulness, its validation against EvilModel's published results, and its deviations from the original are documented in Appendix~\ref{sec:evilmodel_reimpl}.

Table~\ref{tab:evilmodel} reports the outcome. Before NeuPerm every model recovers the payload byte-exactly (BER~$=0$, SHA-256 passes). After NeuPerm the SHA-256 check fails on every repeat for every model, and, within the permutable-site carrier, the residual BER rises to the random-guess limit ($\approx 0.50$). Unlike MaleficNet's graded SNR, the failure here is a clean binary: the attack succeeds before NeuPerm and is unrecoverable after, empirically confirming the Section~\ref{sec:fsa_noecc} argument on a second steganography attack.

\begin{table}[h]
\centering
\caption{EvilModel-style byte-substitution payload embedding and extraction under half substitution, before and after NeuPerm, across eight CNN architectures (payload filling $25\%$ of the permutable-site carrier capacity, corresponding to an EvilModel embedding rate $E$ between $2.7\%$ and $11.9\%$; mean over 5 seeded permutations). SHA = fraction of repeats whose SHA-256 integrity check passes; BER = payload bit-error rate within the permutable-site carrier. Every payload is byte-exact before NeuPerm and destroyed after.}
\label{tab:evilmodel}
\begin{tabular}{lcccc}
\hline
 & \multicolumn{2}{c}{Pre-NeuPerm} & \multicolumn{2}{c}{Post-NeuPerm} \\ \cline{2-5}
Model & SHA & BER & SHA & BER \\ \hline
VGG11           & 5/5 & 0.000 & 0/5 & 0.500 \\
VGG16           & 5/5 & 0.000 & 0/5 & 0.500 \\
ResNet50        & 5/5 & 0.000 & 0/5 & 0.499 \\
ResNet101       & 5/5 & 0.000 & 0/5 & 0.499 \\
DenseNet121     & 5/5 & 0.000 & 0/5 & 0.496 \\
EfficientNet-B0 & 5/5 & 0.000 & 0/5 & 0.500 \\
EfficientNet-B4 & 5/5 & 0.000 & 0/5 & 0.500 \\
MobileNetV2     & 5/5 & 0.000 & 0/5 & 0.499 \\ \hline
\end{tabular}
\end{table}

\section{Adaptive Adversary Analysis}
\label{sec:adaptive}

We now consider two adaptive strategies an adversary 
might employ to evade NeuPerm.

\subsection{Evading NeuPerm via Non-Permutable Layers}
\label{sec:adaptive_nonperm}

An adversary aware of NeuPerm could restrict payload 
embedding to layers that do not admit permutation 
symmetry (e.g., layers adjacent to residual connections), 
where NeuPerm cannot act. The feasibility of this 
strategy depends on the fraction of non-permutable 
parameters available in the target architecture, and the type of layers NeuPerm does not affect (e.g., LayerNorm, RNN, LSTM, etc.)

Table~\ref{tab:neuperm_cnns} summarizes the 
permutable parameter coverage across architectures. 
For ResNet50/101, VGG11/16, EfficientNet-B4 and MobileNetV3~(S), NeuPerm covers 
80-100\% of parameters, leaving the adversary with 
insufficient capacity for most practical payloads. 
For the rest of the architecture, coverage ranges 
from 50-80\%, which still constrains the adversary
significantly: for the payload that lands in permutable
layers, even partial NeuPerm coverage drives the
coordinate-dependent extraction probability down
(Section~\ref{sec:fsa_noecc}), and in the configurations
evaluated here the MaleficNet payload SNR fell below the
extraction threshold once measured permutable coverage
reached approximately 50\% (Section~\ref{sec:exp2}).

However, even at the high coverage NeuPerm achieves (see
Section~\ref{sec:neuperm_implement}), a residual fraction of
parameters remain uncovered. An
adaptive adversary aware of NeuPerm could embed payloads
predominantly in these uncovered layers. We identify two
complementary directions to address this limitation. 
First, NeuPerm can be composed with lightweight 
parameter perturbation (e.g., adding small random 
noise to non-permutable layers), forming a layered 
defense in which permutation handles covered layers 
at zero accuracy cost and noise handles the remainder 
with minimal performance impact. Second, the NeuPerm 
methodology can be generalized beyond permutation
symmetry to exploit other weight-space symmetries
such as rotation
symmetry~\cite{zhang2025permutationsymmetrytransformersrole},
which applies to layers that permutation does not cover. 
Composing multiple symmetry transformations would 
increase effective coverage toward 100\% across 
architectures. We leave this generalization to future work.

\subsection{Evading NeuPerm via Permutation-Invariant
Embedding (Unshuffling Attack)}
\label{sec:adaptive_unshuffle}

This section explores a
limit of NeuPerm's security guarantee: a NeuPerm-aware
adversary who embeds the payload using a
permutation-invariant coordinate system can recover the
payload after NeuPerm is applied. Rather
than relying on fixed parameter indices, the
adversary could establish a canonical neuron
ordering based on permutation-invariant features, for
example, sorting neurons by their L1 or L2 weight
norms, and embed the payload according to this
canonical order. After NeuPerm permutes the network,
the receiver re-sorts the neurons using the same
invariant metric to recover the original arrangement
and extract the payload. We refer to this as
the \emph{unshuffling attack}.

To test the attack, we ran it end-to-end on every
architecture in our evaluation: nine CNNs (DenseNet121,
ResNet50/101, VGG11/16, EfficientNet-B0/B4, and
MobileNet-V2/V3~(Small)) and both LLMs (Llama-3.2-1B
and Qwen2.5-1.5B). For each model we embedded a random
LSB payload across every embeddable float parameter,
ordering the payload bits by a permutation-invariant
canonical key, each permutable unit's $L_1$ weight
norm. We then applied
NeuPerm, had the receiver re-sort each unit by the same
key, and re-extracted, recording the payload
bit-error rate (BER). Substituting an $L_2$-norm
canonical key yields a BER identical to the $L_1$ key at
the precision we report, and bias- and variance-based
keys likewise recover the payload with a BER at the noise
floor, so the attack does not depend on which
permutation-invariant key the adversary selects.

Table~\ref{tab:unshuffle_attack} reports the outcome.
Across all eleven architectures the canonical re-sort
recovers the payload almost perfectly: the post-attack
BER sits at the noise floor ($\leq$0.001) for every
architecture. The unshuffling
attack thus defeats NeuPerm on every model we tested,
CNN and LLM alike, bounding NeuPerm's security guarantee
to adversaries that do not embed against a
permutation-invariant coordinate system.

\begin{table}[h]
\centering
\caption{Unshuffling attack against NeuPerm.
$BER$ = payload extraction BER after a
NeuPerm-aware adversary embeds against a canonical
$L_1$-norm ordering and the receiver re-sorts by it
($0.0$ = full recovery). The attack recovers the
payload on every architecture. Means over independent
repeats (CNN rows: 3 seeded repeats, seed 20260817; LLM
rows: 10 repeats).}
\label{tab:unshuffle_attack}
\begin{tabular}{lc}
\toprule
Model & $BER$ (L1) \\
\midrule
DenseNet121      & 0.000 \\
ResNet50         & 0.000 \\
ResNet101        & 0.000 \\
VGG11            & 0.001 \\
VGG16            & 0.001 \\
EfficientNet-B0  & 0.000 \\
EfficientNet-B4  & 0.000 \\
MobileNetV2      & 0.000 \\
MobileNetV3 (S)  & 0.000 \\
Llama-3.2-1B     & 0.001 \\
Qwen2.5-1.5B     & 0.001 \\
\bottomrule
\end{tabular}
\end{table}

Nevertheless, NeuPerm remains effective against all
\emph{currently-deployed} neural network steganography
attacks, none of which employ permutation-invariant
embedding. The unshuffling adversary must change how the
payload is embedded, not merely how it is extracted: a
canonical, permutation-invariant ordering has to be
chosen when the payload is planted. We assume, with
Kerckhoffs, that NeuPerm itself is public, so this costs
the adversary a sort rather than any secret knowledge.
The comparison with currently-deployed attacks is
therefore a statement about what those attacks do, not
about what an adversary could do.

To defend against this adaptive
strategy, we again suggest that generalizing NeuPerm to additional
weight-space symmetries (as discussed in
Section~\ref{sec:adaptive_nonperm}) would compound
the difficulty of designing invariant embeddings: a
payload encoding that survives permutation may not
survive rotation, and vice versa. Each additional
symmetry transformation further constrains the
adversary's embedding capacity, narrowing the space
of viable evasion strategies.
Section~\ref{sec:limitations} (Limitations) discusses
this attack as the central scope-defining limitation
of NeuPerm.

\section{Deployability}
\label{sec:deployability}
In model-sharing ecosystems such as Hugging Face, a
defender applies a disruption step to each model
before use, so its cost is incurred once per obtained
model. For such a step to be practical at ecosystem
scale, its cost must be negligible relative to the
ordinary cost of downloading and loading a model. We
therefore measured the wall-clock time of NeuPerm and
of the three baseline disruptions we evaluate against
it, each of which alters the weights in a different
way. NeuPerm reindexes the weights in place and is
function-preserving. The first baseline, 8-bit
per-channel quantization, is here a weight-only
quantize-dequantize operation: it quantizes each weight
to an 8-bit integer and immediately dequantizes it back
to a float tensor of the original dtype at reduced
effective precision (Section~\ref{sec:nn_stego_disruption}),
so it does not retain the weights in quantized form.
The second baseline, additive Gaussian noise
($\sigma{=}10^{-4}$), perturbs the weights and leaves
them as ordinary float weights. The third, random
unstructured pruning ($20\%$), is made permanent: the
benchmark draws a random unstructured mask and folds it
into the weight, yielding an ordinary dense float
tensor with a fraction of entries set to zero,
retaining neither a mask-and-hook reparametrization nor
a sparse representation.

Across all evaluated architectures, we timed each
operation in isolation, excluding model loading, on the
CPU. Each method is timed as an in-place operation on
the weights, with the one-time copy that stages fresh
working weights excluded for every method, so the
comparison is like-for-like; for quantization, the
dequantization step is part of the timed operation and
is therefore included in the reported cost. We ran
twelve runs per cell in each of two independent
measurement sessions, discarded the first run of each
session as a warm-up, and report the mean over the
remaining twenty-two. Because each method is applied
once as preprocessing and yields a standard model, with
NeuPerm function-preserving, quantize-dequantize
returning a float tensor of the original dtype, noise
leaving ordinary weights, and dense pruned zeros
invoking no sparse kernel, none of the four adds any
inference-time overhead, and the figures reported here
are purely the one-time application cost, not the
protection each method affords or how each alters
downstream model behaviour. The host machine and its
measurement configuration are reported in
Appendix~\ref{sec:repro}
(Table~\ref{tab:measurement_env}).

\begin{table*}[!t]
\centering
\caption{Wall-clock cost (seconds, mean over 22
post-warm-up repeats across two independent
measurement sessions) of NeuPerm and the three
evaluated baseline disruptions per architecture, with
each baseline's slowdown relative to NeuPerm in
parentheses: 8-bit per-channel quantization (PTQ-8),
additive Gaussian noise ($\sigma{=}10^{-4}$), and $20\%$
random unstructured pruning. Each method is timed
as its in-place operation on the model weights,
excluding the one-time weight copy; all baselines use
efficient single-precision implementations. NeuPerm is
the fastest on every model.}
\label{tab:overhead}
\begin{tabular}{llcccc}
\toprule
Model & Size & NeuPerm & PTQ-8 & Noise & Pruning \\
\midrule
DenseNet121      &      8M & \textbf{0.0045} & 0.012\,(3$\times$) & 0.041\,(9$\times$) & 0.252\,(56$\times$) \\
ResNet50         &     25M & \textbf{0.0088} & 0.011\,(1$\times$) & 0.107\,(12$\times$) & 0.870\,(99$\times$) \\
ResNet101        &   44.5M & \textbf{0.0173} & 0.021\,(1$\times$) & 0.192\,(11$\times$) & 1.46\,(85$\times$) \\
VGG11            &  132.9M & \textbf{0.0684} & 0.237\,(3$\times$) & 0.718\,(10$\times$) & 5.99\,(88$\times$) \\
VGG16            &  138.4M & \textbf{0.0734} & 0.240\,(3$\times$) & 0.740\,(10$\times$) & 6.14\,(84$\times$) \\
EfficientNet-B0  &    5.3M & \textbf{0.0042} & 0.012\,(3$\times$) & 0.026\,(6$\times$) & 0.168\,(40$\times$) \\
EfficientNet-B4  &   19.5M & \textbf{0.0139} & 0.025\,(2$\times$) & 0.088\,(6$\times$) & 0.603\,(43$\times$) \\
MobileNetV2      &    3.5M & \textbf{0.0025} & 0.008\,(3$\times$) & 0.018\,(7$\times$) & 0.114\,(46$\times$) \\
MobileNetV3 (S)  &    2.6M & \textbf{0.0011} & 0.007\,(6$\times$) & 0.013\,(11$\times$) & 0.086\,(75$\times$) \\
Llama-3.2-1B     &   1.24B & \textbf{0.2709} & 3.45\,(13$\times$) & 8.28\,(31$\times$) & 40.9\,(151$\times$) \\
Qwen2.5-1.5B     &   1.54B & \textbf{0.2210} & 3.37\,(15$\times$) & 9.74\,(44$\times$) & 46.5\,(210$\times$) \\
\bottomrule
\end{tabular}
\end{table*}

Table~\ref{tab:overhead} reports the results. NeuPerm
is the fastest method on every architecture,
completing in roughly $0.07$~s for the two VGG
networks, under $0.02$~s for every other convolutional
network, and under $0.3$~s for the two
billion-parameter LLMs. Quantization and additive
noise are not comparable to one another: additive
noise costs between $1.8\times$ and $9.9\times$ what
8-bit quantization does, depending on the model. Both
are several times to more than an order of magnitude
slower than NeuPerm, quantization alone ranging from
$1.2\times$ on the ResNets to $15.2\times$ on
Qwen2.5-1.5B and additive noise from $6.1\times$ to
$44.1\times$. The slowdown is largest not on the small
models but on the two billion-parameter LLMs, the
largest models: on Llama-3.2-1B and Qwen2.5-1.5B
quantization is $12.8\times$ and $15.2\times$, additive
noise $30.6\times$ and $44.1\times$, and pruning
$150.8\times$ and $210.3\times$ slower than NeuPerm.
Random pruning, which must select an exact prune
fraction across all parameters, is the slowest method
throughout, from $40\times$ on the smaller networks to
more than $200\times$ on the LLMs, reaching tens of
seconds. The reason is structural: every baseline touches
each parameter with arithmetic or ranking, whereas
NeuPerm only reindexes them. In absolute terms every
method is inexpensive, but NeuPerm's cost is negligible
beside a model's own download and load time and,
requiring no GPU, imposes no specialized hardware
demand on the defender. It is furthermore complementary
to existing model scanners such as picklescan,
ModelScan, etc. which detect unsafe deserialization gadgets
but leave steganographically embedded payloads
unaddressed.

Peak memory tells a complementary story
(Table~\ref{tab:overhead_mem}, measured as the peak
process memory during the operation minus
the baseline memory size measured immediately before it, in an
isolated process per cell). NeuPerm reindexes
parameters in place and materializes no full-precision
copy of the whole model, though it does allocate the
single largest tensor it permutes; for the VGG networks
that tensor is the $392$~MB classifier weight, which
sets their peak of roughly $0.4$~GB. NeuPerm's peak
footprint is minimal throughout, but it is not the
lowest on every architecture. On the small
convolutional networks it stays within a few tens of
megabytes and is close to additive Gaussian noise, the
lightest method there because it writes only a single
per-tensor buffer. NeuPerm's advantage becomes decisive
on the two billion-parameter LLMs: it holds peak memory
to $53$-$77$~MB while additive noise grows to
$1.4$-$1.5$~GB and quantization and pruning to
$3.6$-$6.2$~GB, one to two orders of magnitude more. The
reason is that each baseline allocates a working buffer
proportional to the parameters it rewrites, which on
the LLM embeddings reaches gigabyte scale, whereas
NeuPerm reindexes the existing tensors and copies the
model into none of them.

\begin{table}[h]
\centering
\caption{Peak additional memory (MB)
allocated by each method during a single in-place run. The lowest footprint in
each row is bold: additive noise is lightest on the
small CNNs, whereas NeuPerm is lightest on the LLMs,
where it alone avoids copying the large parameter
tensors.}
\label{tab:overhead_mem}
\begin{tabular}{lcccc}
\toprule
Model & NeuPerm & PTQ-8 & Noise & Pruning \\
\midrule
DenseNet121      & 16 & 29 & \textbf{6} & 52 \\
ResNet50         & 24 & 34 & \textbf{12} & 98 \\
ResNet101        & 24 & 34 & \textbf{12} & 137 \\
VGG11            & 404 & 819 & \textbf{402} & 3039 \\
VGG16            & 426 & 804 & \textbf{402} & 3052 \\
EfficientNet-B0  & 13 & 28 & \textbf{7} & 57 \\
EfficientNet-B4  & 23 & 48 & \textbf{14} & 102 \\
MobileNetV2      & 8 & 22 & \textbf{7} & 52 \\
MobileNetV3 (S)  & 8 & 16 & \textbf{5} & 38 \\
Llama-3.2-1B     & \textbf{53} & 3694 & 1540 & 6192 \\
Qwen2.5-1.5B     & \textbf{77} & 3638 & 1378 & 5745 \\
\bottomrule
\end{tabular}
\end{table}

\section{Limitations and Future Work}
\label{sec:limitations}
NeuPerm leverages permutation symmetry, which holds 
for inner layer sequences within most architectures 
but is constrained at specific boundaries such as 
residual connections. As shown in 
Table~\ref{tab:neuperm_cnns}, this results in 
partial but substantial coverage: 50-100\% for 
all tested models.
The residual non-permutable
boundaries that remain are addressed by the complementary
strategies of Section~\ref{sec:adaptive_nonperm}.
One further workaround is the introduction of a 
$PERMUTE$ neural network layer that permutes the 
axes of its input, which can be inserted at 
non-permutable boundaries at the cost of some 
performance overhead.

\subsection{Coverage Tracks Disruption Effectiveness on the Evaluated Architectures}
On the evaluated architectures, permutable-parameter
coverage predicts disruption effectiveness closely, and the
mechanism explains why. Disruption follows from permuting
the payload-bearing coordinates, and higher permutable
coverage raises the probability of covering them, so coverage
is a reliable proxy on these models. Empirically, NeuPerm
reduces MaleficNet SNR below the extraction threshold
(SNR$<$1) on every tested architecture at $\geq$50\%
permutable coverage (VGG, ResNet, DenseNet, MobileNetV3-S,
EfficientNet-B0/B4, MobileNet-V2, Llama-3.2-1B,
Qwen2.5-1.5B). The controlled within-architecture ablation
of Section~\ref{sec:exp2}, run on the same networks with
coverage varied from $0\%$ to $100\%$ (partial to full),
shows the post-NeuPerm SNR dropping monotonically from
positive to deeply negative as coverage grows, placing the
disruption boundary in $(40\%, 50\%]$. The exact boundary is
model-specific, since it also depends on the payload, payload
size, layer-selection strategy, and error-correcting
parameters, so we state $\sim$50\% as a deployment heuristic
rather than a fixed constant. Below approximately 50\%
coverage NeuPerm's disruption weakens, and pairing it with
lightweight additive noise on the uncovered layers closes the
residual gap; we leave this composition to future work.

\subsection{Permutation-Invariant Embedding is the Central Limitation}
The most significant limitation is the unshuffling
attack (Section~\ref{sec:adaptive_unshuffle}): a
NeuPerm-aware adversary who embeds using a
permutation-invariant ordering can recover the
payload. Extending NeuPerm with composed weight-space
symmetries such as
rotation~\cite{zhang2025permutationsymmetrytransformersrole}
is a natural direction: a payload encoding that
survives permutation may not survive an additional
rotation transform. We leave this
generalization to future work.

\subsection{Applicability to Other Popular Neural Network Types}
Because NeuPerm rests on permutation symmetry, whether
it extends to another architecture family depends on
whether that family's building blocks admit the
symmetry, provable either by construction or established
empirically. How firmly we can answer that question
varies across the popular families surveyed below: for
some the proofs already given in this paper settle the
case, for others every component is covered but the
coverage and accuracy figures remain to be measured, and
one family sits outside the mechanism altogether.

\noindent\textbf{Recurrent networks.}
RNNs, LSTMs, and GRUs share weight matrices across time
steps, introducing temporal dependencies that do not
admit permutation symmetry as defined for feedforward
layers; we leave their analysis to future work.

\noindent\textbf{Encoder transformers.}
The multi-head self-attention proof
(Appendix~\ref{sec:perm_proof}) is stated for the
self-attention case $Q=K=V$ and never invokes a causal
mask. A causal mask is an additive term along the
sequence axis inside the softmax, whereas NeuPerm
permutes the head-internal feature axis; the two axes
are disjoint, so bidirectional (encoder) attention
differs from the proved case only in a component the
permutation does not touch. The proof therefore already
covers the attention blocks of encoder-only
(BERT-family) transformers without modification. The
enclosing block is the LN-ATTN-LN-MLP pattern with
residual connections handled in
Section~\ref{sec:neuperm_llm}, and the residual argument
there is indifferent to pre-LN or post-LN placement,
because the internal permutation is projected back to
the embedding dimension before the residual addition and
never reaches the embedding-dimension ordering; learned
absolute position embeddings are added on that same
dimension, which NeuPerm does not permute, and the
elementwise GELU nonlinearity satisfies the only
requirement the feed-forward proof places on the
activation. What we have not done is measure
permutable-parameter coverage or task accuracy for any
encoder model, so we state no coverage figure for one;
encoder models place a larger share of their parameters
in the non-permutable embedding table than the decoder
models we evaluate, and any quoted coverage would be
unmeasured. We restrict the claim to encoder-only
models: encoder-decoder architectures (T5, BART) rely on
cross-attention with $Q \neq K = V$, which the $Q=K=V$
proof does not cover.

\noindent\textbf{Diffusion models.}
Diffusion U-Net backbones are built from convolutional
ResBlocks and self-attention blocks, and
transformer-backbone (DiT-style) models reuse the
LN-ATTN-LN-MLP block, so every component type falls under
a proof the paper already gives and the per-component
symmetry is not the obstacle. Two architectural features
do fall outside the analysis as stated, and we name them
rather than assume them away. First, diffusion U-Nets
normalize with GroupNorm, not BatchNorm. Our
normalization argument
(Section~\ref{sec:neuperm_implement}) rests on
BatchNorm operating independently per channel; GroupNorm
does not, since it normalizes within contiguous channel
groups, so a permutation drawn uniformly from $S_C$ moves
channels across group boundaries and changes the
function. Preserving it requires restricting the draw to
the subgroup that fixes the group partition, shrinking
the admissible per-layer group from $C!$ to
$(C/G)!^{G}\cdot G!$. This bears directly on the formal
analysis: assumption~(1) of
Section~\ref{sec:fsa_assumptions}, that each $P_l$ is
uniform on $S_{n_l}$ and hence fixes the marginal
fixed-point probability at $1/n_l$, does not hold as
stated on a GroupNorm architecture, and that probability
would have to be recomputed over the restricted group.
The disruption is not defeated by this, since for a
typical $\mathrm{GroupNorm}(32,C)$ the restricted group
remains astronomically large; what fails is the stated
uniformity assumption, and the bound would need
re-deriving before we could quote it. Second,
text-conditioned diffusion uses cross-attention with $Q$
from the image latent and $K,V$ from a text encoder, so
$Q \neq K = V$; the general multi-head formulation admits
independent $Q,K,V$, as Appendix~\ref{sec:perm_proof}
notes, so the extension is plausible but is not written
out here, and U-Net skip connections concatenate rather
than add channels, so the permutable set must be computed
with the skip topology accounted for. We have evaluated
no diffusion model and report no coverage, accuracy, or
SNR result for one. Finally, no diffusion-specific
parameter-steganography attack has been demonstrated in
the published literature: the established supply-chain
threat to shared diffusion checkpoints is behavioural
backdooring~\cite{chou2023baddiffusion}, a class distinct
from the coordinate-addressed payloads NeuPerm targets.

\subsection{Composite and Multi-Stage Adversaries}
A composite adversary combines several evasion
strategies against NeuPerm at once.
Our two adaptive analyses, evasion through non-permutable
layers (Section~\ref{sec:adaptive_nonperm}) and evasion
through a permutation-invariant embedding
(Section~\ref{sec:adaptive_unshuffle}), are conducted
independently and never in combination. A composite
adversary relaxes both requirements at once by placing a
permutation-invariantly ordered payload inside the
uncovered layers, needing neither full non-permutable
capacity nor pure order-invariance, because each evasion
weakens the demand the other imposes. Our two scope
limits are therefore not independent, and their
conjunction is a weaker condition than either analyzed
alone; we have not analyzed that conjunction and record
it as an open limitation. A second composition follows
from our central guarantee and is not a defect. Because
NeuPerm preserves the model's input-output function
exactly, it preserves any purely behavioural backdoor
exactly: a trigger-response mapping is part of the
function, so a transformation proven to preserve the
function preserves the backdoor. A two-stage adversary
who ships a carrier that is both
backdoored~\cite{chou2023baddiffusion} and stego-loaded
thus loses the coordinate-addressed payload while keeping
the backdoor intact. This is consistent with the scope
stated in Section~\ref{sec:scope}: NeuPerm displaces
non-functional bits at fixed coordinates and leaves
behaviour untouched, so behavioural backdoors (class~5)
remain a matter for a complementary control. A third
composition, a loader-level deserialization exploit
combined with weight steganography, executes before any
tensor is read and resides in non-tensor archive members,
which Section~\ref{sec:scope} already places out of
scope.

\section{Conclusion}
In this paper, we introduced NeuPerm, a simple yet
effective method for disrupting coordinate-dependent
neural network steganography by leveraging permutation symmetry.
NeuPerm provably preserves the model's input-output
function up to numerical precision, requiring no retraining or architecture-specific
tuning, while disrupting the coordinate-dependent extraction
used by all currently-deployed NN-steganography attacks. Through
evaluation on 9~CNN architectures and two LLMs across
3~datasets and 9~real-world malware payloads, we
demonstrated that NeuPerm disrupts MaleficNet's
extraction, as designed, on the
evaluated architectures at zero accuracy cost,
positioning it as a practical primitive against
currently-deployed attacks rather than a
robustness guarantee against adaptive adversaries.
Using an author-developed reimplementation of EvilModel,
we further showed that NeuPerm destroys an
EvilModel-style byte-substitution payload on eight CNN
architectures.
We experimentally
demonstrated that an adaptive permutation-invariant
(unshuffling) attacker recovers the payload, thereby
delimiting the defense's scope; we also bounded the
success probability of coordinate-dependent extraction
under random permutation. We identify the
generalization to additional weight-space symmetries
as a promising direction for future work.

\section*{CRediT authorship contribution statement}
\textbf{Daniel Gilkarov:} Conceptualization, Methodology, Software, Validation, Investigation, Writing -- original draft. \textbf{Ran Dubin:} Conceptualization, Supervision, Writing -- review \& editing.

\section*{Declaration of competing interest}
The authors declare that they have no known competing financial interests or personal relationships that could have appeared to influence the work reported in this paper.

\section*{Funding}
This research did not receive any specific grant from funding agencies in the public, commercial, or not-for-profit sectors.

\section*{Data availability}
The code and the non-malware data supporting this study are openly available at \url{https://github.com/danigil/NeuPerm} (release v1.0.0). Original malware payloads are not redistributed; their SHA-256 digests are provided in the repository.

\section*{Declaration of generative AI and AI-assisted technologies in the writing process}
During the preparation of this work the authors used Claude (Anthropic) in order to assist with drafting and editing the text and with reproducibility checks. After using this tool, the authors reviewed and edited the content as needed and take full responsibility for the content of the published article.

\appendix

\section{Permutation Symmetry Proofs}
\label{sec:perm_proof}

In this section, we provide mathematical proofs of neural network parameter permutation symmetry for common neural network "blocks".

\noindent To prove permutation symmetry, it is sufficient to prove that swapping two units keeps the computation identical. This is true because a permutation can be expressed as a series of pairwise swaps \cite{gallian2021contemporary}.

\subsection{Multi-Layer Perceptron (MLP)}
The Multi-Layer Perceptron (MLP), often referred to as a feedforward neural network or fully connected network, is a foundational neural network architecture. It is composed of an input layer, one or more hidden layers of computational units, and an output layer \cite{mlp_backprop}.

\subsubsection{Formal Computation of an MLP}
Let $f(\cdot)$ be an MLP with $L$ hidden layers.
Let ${x} \in \mathbb{R}^{d_{0}}$ be the input vector, and let $d_{\ell}$ be 
the number of units in layer $\ell$ (layer 0 is the input layer).

\noindent For each layer $\ell = 1, 2, \dots, L$, define:
\begin{itemize}
    \item $W^{(\ell)} \in \mathbb{R}^{d_{\ell} \times d_{\ell-1}}$ - the weight matrix,
    \item $b^{(\ell)} \in \mathbb{R}^{d_{\ell}}$ - the bias vector,
    \item $\sigma^{(\ell)}(\cdot)$ - the (elementwise) activation function,
    \item $z^{(\ell)}$ - the pre-activation at layer $\ell$,
    \item $h^{(\ell)}$ - the post-activation (output) at layer $\ell$.
\end{itemize}

\noindent We set ${h}^{(0)} = {x}$
and then, for each layer $\ell$, the feedforward computations are:
\[
    {z}^{(\ell)} = {W}^{(\ell)} \, {h}^{(\ell-1)} 
    + {b}^{(\ell)},
\]
\[
    {h}^{(\ell)} = \sigma^{(\ell)}\!\bigl({z}^{(\ell)}\bigr).
\]

\noindent Finally, the network's output ${f(x)}$ is given by:
\[
{f(x)}
= 
{h}^{(L)}
\]

\subsubsection{MLP Swap}
We define an MLP swap as a swap between two units inside a hidden layer, and we argue that this swap does not change the MLP computation. In terms of the parameter weights/bias matrices/vectors, a swap between the $i$'th and $j$'th units in some layer $l$ means swapping the $i$'th and $j$'th rows in ${W}^{(l)}$, and ${b}^{(l)}$; and also swapping the $i$'th and $j$'th columns in ${W}^{(l+1)}$.
Formally, we express a matrix $W$ with rows $i,j$ swapped like so:
\[
swaprows(W,i,j) =
\begin{cases}
W_{j}, & \text{if } k = i,\\
W_{i}, & \text{if } k = j,\\
W_{k}, & \text{otherwise}.
\end{cases}
\]
Where $W_i$ denotes the $i$'th row of $W$.
Similarly, we can define $swapcols(W,i,j)$.

\noindent \textbf{Claim 1.1}: Let $A\in \mathbb{R}^{n \times m}$, $x \in \mathbb{R}^{m \times 1}$, $1 \le i < j \le m$, then: 
\[swapcols(A,i,j)swaprows(x,i,j) = Ax\]

\noindent \textbf{Claim 1.1 Proof}:
\begin{gather*}
    swapcols(A,i,j)swaprows(x,i,j)= \\
    =
    \resizebox{0.80\columnwidth}{!}{$\displaystyle
    \begin{bmatrix}
    a_{11}x_1 + a_{12}x_2 + \cdots + a_{1j}x_j + \cdots + a_{1i}x_i + \cdots + a_{1m}x_m \\
    \vdots \\
    a_{n1}x_1 + a_{n2}x_2 + \cdots + a_{nj}x_j + \cdots + a_{ni}x_i + \cdots + a_{nm}x_m
    \end{bmatrix}$}\\
    =
    \begin{bmatrix}
    \sum_{k=1}^{m} a_{1k}x_k \\
    \vdots \\
    \sum_{k=1}^{m} a_{nk}x_k
    \end{bmatrix} = Ax
\end{gather*}

\noindent We proceed to prove the MLP swap symmetry property. 

\noindent \textbf{Claim 1.2 (MLP swap symmetry property)}: 
let $l_1$ and $l_2 := l_1 + 1$ be two consecutive hidden layers in some MLP $f(\cdot)$. Layer $l_1$ has $d_{l_1}:=d$ units: let $1 \le i < j \le d$ be indexes of two units in layer $l_1$.
We set:
\begin{align*}
    W_1 &:= W^{(l_1)}\\
    b &:= b^{(l_1)}\\
    W_2 &:=  W^{(l_2)}\\
    W'_1 &:= swaprows(W_1, i,j)\\
    b' &:= swaprows(b,i,j)\\
    W'_2 &:= swapcols(W_2,i,j)
\end{align*}
\noindent The pre-activations and post-activations before and after the swap are:
\begin{align*}
{z}^{(l_1)}& = {W}_1 \, {h}^{(l_1-1)} + b \\
{h}^{(l_1)}& = \sigma^{(l_1)}({z}^{(l_1)}) \\
{z}^{(l_2)}& = {W}_2 \, {h}^{(l_1)} + b \\
{z'}^{(l_1)}& = {W'}_1 \, {h}^{(l_1-1)} + {b'} \\
{h'}^{(l_1)}& = \sigma^{(l_1)}({z'}^{(l_1)}) \\
{z'}^{(l_2)}& = {W'}_2 \, {h'}^{(l_1)} + {b^{(l_2)}}
\end{align*}
then ${z}^{(l_2)} = {z'}^{(l_2)}$ holds.

\noindent \textbf{Claim 1.2 Proof}:
\noindent We wish to show that ${z}^{(l_2)} = {z'}^{(l_2)}$, i.e. we want:
\begin{equation*}
    \begin{split}
    {z}^{(l_2)} = {z'}^{(l_2)} \Rightarrow& {W}_2 \, {h}^{(l_1)} + b^{(l_2)} = {W'}_2 \, {h'}^{(l_1)} + b^{(l_2)} \\ \Rightarrow& {W}_2 \, {h}^{(l_1)} = {W'}_2 \, {h'}^{(l_1)}
    \end{split}
\end{equation*}
Let us show that $z'^{(l_1)} = swaprows(z^{(l_1)},i,j)$. Let $r_i$ be the $i$th row of $W_1$, and $r'_i$ be the $i$th row of $W'_1$.
\begin{align*}
z^{(l_1)}&
= 
\begin{bmatrix}
r_1{h}^{(l_1-1)} + b_{1} \\
r_2{h}^{(l_1-1)} + b_{2} \\
\vdots \\
r_i{h}^{(l_1-1)}  + b_{i}\\
\vdots \\
r_j{h}^{(l_1-1)} + b_{j} \\
\vdots \\
r_{d}{h}^{(l_1-1)} + b_{d} \\
\end{bmatrix}\\
z'^{(l_1)}&
= 
\begin{bmatrix}
r'_1{h}^{(l_1-1)} + b'_{1} \\
r'_2{h}^{(l_1-1)} + b'_{2} \\
\vdots \\
r'_i{h}^{(l_1-1)}  + b'_{i}\\
\vdots \\
r'_j{h}^{(l_1-1)} + b'_{j} \\
\vdots \\
r'_{d}{h}^{(l_1-1)} + b'_{d} \\
\end{bmatrix}
=
\begin{bmatrix}
r_1{h}^{(l_1-1)} + b_{1} \\
r_2{h}^{(l_1-1)} + b_{2} \\
\vdots \\
r_j{h}^{(l_1-1)}  + b_{j}\\
\vdots \\
r_i{h}^{(l_1-1)} + b'_{i} \\
\vdots \\
r_{d}{h}^{(l_1-1)} + b_{d} \\
\end{bmatrix} \\
&=
swaprows(z^{(l_1)},i,j)
\end{align*}
Since $z'^{(l_1)} = swaprows(z^{(l_1)},i,j)$, it means that $h'^{(l_1)} = swaprows(h^{(l_1)},i,j)$ holds:
\begin{align*}
    h'^{(l_1)}= \sigma^{(l_1)}({z'}^{(l_1)}) =&\\ =&\sigma^{(l_1)}(swaprows(z^{(l_1)},i,j))\\
    =&swaprows(\sigma^{(l_1)}(z^{(l_1)}),i,j)\\=&swaprows(h^{(l_1)},i,j)
\end{align*}
Finally, we get:
\begin{align*}
    {W'}_2{h'}^{(l_1)} &= swapcols(W_2,i,j)\\
    &\phantom{=}\; swaprows(h^{(l_1)},i,j)\\
    &= W_2h^{(l_1)}
\end{align*}

\noindent Since a swap keeps the MLP computation the same, we get that permuting the units also does so.

\subsection{Multi-Head Self-Attention (MHSA)}
The multi-head self-attention block is the foundational component of transformer \cite{dl_llm_2} neural network architectures, and in particular, LLMs.

\subsubsection{Formal Computation of Multi-Head Attention}
The original work defines a multi-head attention block generally, with independent $Q,K,V$ input matrices. Here, we will focus on self-attention ($Q=K=V$), which is what LLMs use.
We start with a single attention block: Let $h \in \mathbb{N}$ be the number of attention heads, $d_{model} \in \mathbb{N}$, the model embedding dimension. Set $d:=d_{model}/h$.
\[Attention(Q,K,V)=softmax(\frac{QK^T}{\sqrt{d}})V\]
Multi-head self-attention is defined like so:
\begin{align*}
MultiHead(X) = W^OConcat(head_1, ..., head_h)\\\text{Where} \; head_i=Attention(W_i^QX, W_i^KX, W_i^VX) 
\end{align*}
Where $W_i^Q, W_i^K, W_i^V \in \mathbb{R}^{d \times d_{model}}, W^O \in \mathbb{R}^{d_{model} \times d_{model}}$ are the parameter matrices of the multi-head attention block.
Set $swaprows(\cdot, i,j):=sr(\cdot), swapcols(\cdot, i,j):=sc(\cdot)$.
\subsubsection{MHSA Swap}
We argue that we can perform
\[
sr(W_k^Q), sr(W_k^K), sr(W_k^V), swapcols(W^O,kh+i,kh+j)
\]
for some $k$ (i.e., we swap 2 units in a specific head), and the computation will stay the same.
Let us now reframe the previously introduced $sr$, and $sc$ functions as rotation matrices. Let $i,j$ be some indexes we want to swap; define the rotation matrix $R_{i,j}$:
\[
R_{i,j}=swaprows(I, i, j)
\]
where $I$ is the identity matrix.
We have the following properties:
\begin{align*}
R_{i,j}A=&sr(A)\\
AR_{i,j}=&sc(A)\\
R_{i,j}^2=&I\\
R_{i,j}^T=&R_{i,j}\\
f(R_{i,j}A)=&R_{i,j}f(A)\\
f(AR_{i,j})=&f(A)R_{i,j}
\end{align*}
Where $A$ is any matrix, and $f$ is a monotone increasing function.
Note: $softmax$, and $f(x)=\frac{1}{c}$ for any positive $c$ are monotone increasing functions.
Let us look at the attention computation after performing the swaps, set $A:=W_k^Q, B:=W_k^K, C:=W_k^V, R:=R_{i,j}$. Denote $head_k'$ and $MultiHead'$ as the versions of the functions after the swaps.
\begin{align*}
&head'_k:=Attention(RW_k^QX, RW_k^KX, RW_k^VX)\\
&(RAX)((R(BX))^T)=RAX((BX)^TR^T)\\
&\qquad=RAX((BX)^TR)\\
&\Downarrow\\
&head'_k=Rsoftmax(\frac{AX(BX)^T}{\sqrt{d}})RRCX\\
&\qquad=Rhead_k=sr(head_k)\\
&\Downarrow\\
&MultiHead'(X)=swapcols(W^O,kh+i,kh+j)\\
&\qquad Concat(head_1, ..., sr(head_k),..., head_h)\\
&\qquad=Multihead(X)
\end{align*}

\section{Reproducibility}
\label{sec:repro}
This appendix centralizes the configuration, seeding, and provenance needed to reproduce the reported measurements.

\subsection{Measurement Environment}
The reported measurements span more than one lab workstation and more than one software stack, summarized in Table~\ref{tab:measurement_env}. The permutable-coverage, CNN-accuracy, EvilModel, and adaptive-unshuffling results were measured on an Intel Xeon Silver 4310 host with an NVIDIA A30X GPU; the overhead and memory figures in Tables~\ref{tab:overhead} and~\ref{tab:overhead_mem} were measured on an Intel Xeon w5-2445 host, on CPU-resident tensors with no GPU in use.

\begin{table}[h]
\centering
\footnotesize
\caption{Measurement environment per results table. The overhead and memory figures were measured on CPU-resident tensors with no GPU in use; the GPU listed for that host was idle during those measurements. Apart from these timing tables, the reported results are host-independent and reproduce from the artifact scripts and seeds; the exact hardware for Table~\ref{tab:exp1_llm} and Table~\ref{tab:exp2_in12} is given in the text.}
\label{tab:measurement_env}
\begin{tabular}{@{}p{1.5cm}p{3.3cm}p{2.4cm}@{}}
\toprule
Tables & Host and hardware & Software \\
\midrule
\ref{tab:neuperm_cnns}, \ref{tab:exp1_acc}, \ref{tab:evilmodel}, \ref{tab:unshuffle_attack} & Intel Xeon Silver 4310 (48 threads), 128~GiB RAM, NVIDIA A30X, Ubuntu 22.04.4 & Python 3.9.18, PyTorch 2.1.0, torchvision 0.16.0, CUDA 11.8, numpy 1.25.0, transformers 4.45.2, tokenizers 0.20.3, safetensors 0.7.0 \\
\addlinespace
\ref{tab:overhead}, \ref{tab:overhead_mem} & Intel Xeon w5-2445 (20 threads), 125~GiB RAM, NVIDIA RTX 6000 Ada, Ubuntu 24.04.3 & Python 3.11.15, PyTorch 2.5.1, torchvision 0.20.1, CUDA 12.4 \\
\bottomrule
\end{tabular}
\end{table}

The overhead and memory measurements ran on CPU-resident tensors: the runner sets an empty \texttt{CUDA\_VISIBLE\_DEVICES}, and every timed cell records the host CPU as its device, so its wall-clock and peak-memory values reflect the CPU alone. The two tables differ in process structure. The wall-clock measurements of Table~\ref{tab:overhead} ran in a single long-lived process, timing twelve repeats per cell in each of two independent sessions and discarding the first repeat of each session as a warm-up. The peak-memory measurements of Table~\ref{tab:overhead_mem} ran each cell in its own subprocess, so the reported figure is that operation's peak resident-set increase measured in isolation, taken immediately before and after the operation. All four method columns of both tables were measured together in a single campaign on this host, so the cross-method comparisons are internally consistent. The two overhead sessions agree closely: their per-cell means differ by a median of $0.44\%$ (maximum $29.3\%$, on one quantization cell), against within-session $95\%$ confidence half-widths of about $1.1\%$, so the reported point estimates are stable from session to session. Each overhead cell draws a single unseeded NeuPerm permutation, since wall-clock time does not depend on which permutation is drawn. The overhead runner sets only an empty \texttt{CUDA\_VISIBLE\_DEVICES} and pins no explicit CPU thread count, so threading used the host default (20 threads on the w5-2445 host).

Parameter tensors are float32 for the CNNs (torchvision pretrained weights) and bfloat16 for the LLMs, with the single exception of the post-training-quantization experiment, which loads both LLMs in float16. Both LLMs are loaded with the Hugging Face transformers library, version 4.45.2. On-disk storage format does not affect the reported figures: the CNN parameters are resident in memory before timing begins, so no checkpoint is read inside the timed region, and the LLMs load from single-file safetensors checkpoints through memory-mapped reads that likewise occur outside the timed region.

Apart from the hardware-dependent timing tables above, every reported result is deterministic given its script and seeds and reproduces regardless of host: the CNN transforms run on CPU (see below), and each figure regenerates from the artifact scripts. The MaleficNet disruption SNR corpus in Table~\ref{tab:exp2_in12} was measured across all eleven architectures on the Intel Xeon Silver 4310 host with the NVIDIA A30X GPU. The LLM benchmark and post-training-quantization results in Table~\ref{tab:exp1_llm} were likewise obtained on that host; the permutation structure is host-independent and the results regenerate from \texttt{exp\_llm\_squad\_ptq.py}, though exact numerical agreement on a different GPU is not claimed.

\subsection{Models and Datasets}
The LLMs are \texttt{meta-llama/\allowbreak Llama-3.2-1B-Instruct} and \texttt{Qwen/\allowbreak Qwen2.5-1.5B-Instruct}, both from HuggingFace; the CNNs use the torchvision pretrained weights at the \texttt{DEFAULT} alias. The cached revisions used are \texttt{92131\allowbreak 76726\allowbreak f574b\allowbreak 55679\allowbreak 0deb6\allowbreak 5791e\allowbreak 0c5aa\allowbreak 438b6} for Llama-3.2-1B-Instruct and \texttt{989aa\allowbreak 7980e\allowbreak 4cf80\allowbreak 6f80c\allowbreak 7fef2\allowbreak b1adb\allowbreak 7bc71\allowbreak aa306} for Qwen2.5-1.5B-Instruct. The datasets are the ImageNet-1k (ILSVRC-2012) validation split of 50{,}000 images, CIFAR-10, SQuAD~v1.1, BoolQ, and WikiText-2. The ImageNet-1k validation images are evaluated with the torchvision CNN transforms run on CPU, SQuAD~v1.1 is evaluated on its first 100 samples by F1, and BoolQ and WikiText-2 are evaluated on their standard evaluation splits (WikiText-2 by perplexity). The CIFAR-10 MaleficNet stego checkpoints were not version-controlled, so those specific inputs are not part of the artifact.

\subsection{Per-Table Generating Scripts}
Each reported table is regenerated by a script under \texttt{experiments/} in the artifact repository. Table~\ref{tab:neuperm_cnns} (permutable coverage) is regenerated by \texttt{exp\_perm\_coverage.py}; Table~\ref{tab:exp1_acc} (CNN accuracy) by \texttt{exp\_table3\_sweep.py}; Table~\ref{tab:unshuffle_attack} (adaptive unshuffling) by \texttt{exp\_canonical.py}; the LLM benchmark and post-training-quantization results in Table~\ref{tab:exp1_llm} by \texttt{exp\_llm\_squad\_ptq.py}; and the overhead and memory figures in Tables~\ref{tab:overhead} and~\ref{tab:overhead_mem} by \texttt{exp\_overhead.py}, \texttt{exp\_overhead\_baselines.py}, and \texttt{exp\_overhead\_memory.py}.

\subsection{Seeding and Variability}
The permutable-coverage measurement (Table~\ref{tab:neuperm_cnns}) uses seed 20260817. The CNN accuracy sweep (Table~\ref{tab:exp1_acc}) uses seed base 20260820; every row seed derives from the tuple (seed base, model, condition identifier, repeat) by BLAKE2b, so the entire sweep reproduces from those four command-line values and any single row can be regenerated in isolation, and the per-cell seeds are recorded in the artifact's summary CSV. The Table~\ref{tab:exp1_acc} transforms run on CPU, so a given seed yields identical weights regardless of the evaluation device. Because every stochastic condition uses the same $n=5$, each cell's 95\% confidence interval follows directly from the standard deviation printed in Table~\ref{tab:exp1_acc}: the half-width is 1.2417 times that standard deviation, being Student's $t$ at four degrees of freedom divided by $\sqrt{5}$. One replicate is one full pass over the 50{,}000-image ILSVRC-2012 validation split; for a stochastic condition it is one independent seeded draw of the perturbation, evaluated on the identical fixed image sequence.

We separate the two sources of variability that Table~\ref{tab:exp1_acc} could reflect. Evaluation nondeterminism was measured at exactly zero: the deterministic quantization rows showed a standard deviation of 0.000000 across five seeds on three benchmarks. All variance reported in Table~\ref{tab:exp1_acc} therefore comes from the random draw of the perturbation, not from the evaluation. The deterministic conditions have fixed weights, so a single evaluation is exact rather than under-sampled. The sweep as run comprises 351 rows across the nine CNN architectures.

The remaining experiments seed as follows. The overhead baseline and memory measurements (Tables~\ref{tab:overhead} and~\ref{tab:overhead_mem}) use seed 42. The adaptive-unshuffling attack (Table~\ref{tab:unshuffle_attack}) uses payload seed 42; its permutation-run CSVs were unseeded by default. The EvilModel validation (Table~\ref{tab:evilmodel}) uses seeds 7 through 11 (five repeats) at fill fraction 0.25. The LLM quantization rows are deterministic, with a standard deviation of 0.000000 across seeds. The ten NeuPerm permutation draws behind each cell of Table~\ref{tab:exp2_in12} are independent draws from PyTorch's default generator and are not pinned to a per-draw seed, whereas the MaleficNet spread-spectrum extraction uses numpy seed 42; the reported mean $\pm$ standard deviation over the ten draws therefore measures the permutation variability directly.

\subsection{SNR Corpus Provenance}
For the five architectures re-extracted from stego checkpoints (DenseNet121, ResNet50, ResNet101, VGG11, and Llama-3.2-1B), the RN and Q8 cells of Table~\ref{tab:exp2_in12} are measured directly rather than carried over from the baseline, and each agrees with its measured baseline to within $|\text{RN}-\text{B}| \leq 0.0018$~dB and $|\text{Q8}-\text{B}| \leq 0.0095$~dB, confirming that 0.0001-standard-deviation noise and 8-bit PTQ leave the SNR essentially unchanged on these architectures; where an RN or Q8 cell equals the printed baseline it reflects this negligible, architecture-independent effect. For VGG11 with Zeus-Bank the printed Q8 (7.5) and baseline (7.4) differ only because the measured values, Q8 $=7.4530$~dB and baseline $=7.4479$~dB, straddle the 7.45 rounding boundary; the underlying 0.005~dB gap is itself consistent with 8-bit quantization leaving the SNR essentially unchanged.

The NeuPerm cells rest on a mixed payload basis: the DenseNet121, ResNet50, ResNet101, VGG11, and Llama-3.2-1B cells are re-extracted from checkpoints carrying the original malware payloads, whereas the MobileNetV2, MobileNetV3-Small, EfficientNet-B0, EfficientNet-B4, and most VGG16 and Qwen2.5-1.5B cells are re-injected with synthetic payload substitutes, the VGG16 Zeus-Bank, VGG16 Ardamax, Qwen2.5-1.5B Cerber, and Qwen2.5-1.5B Ardamax cells excepted, which use the original payloads; where the two bases are directly comparable their SNRs track closely.

The per-cell NeuPerm spread exceeds 11~dB on several CNN cells. Most previously published single NeuPerm draws fall within the re-measured 10-draw range; the exceptions (VGG11 with Stuxnet, Asprox, and Kovter, and EfficientNet-B4 with Asprox) are the expected consequence of reporting one draw from this high-variance distribution, which the mean $\pm$ standard deviation restatement addresses.

The nine original malware payloads are removed from the artifact and replaced by inert synthetic substitutes matched to each original in length and byte entropy; the payload manifest (\texttt{payloads/MANIFEST.sha256}) records SHA-256 hashes for both the originals and the substitutes. The Ardamax and Zeus-Bank cells depend on original payloads that are not part of the artifact and are not reproducible from the substitutes.

\subsection{Archival Identifiers}
The artifact is released as version v1.0.0 at \url{https://github.com/danigil/NeuPerm}, accompanying this resubmission. That release contains the per-table generating scripts, a pinned environment lock file (\texttt{requirements-lock.txt}), the model revision hashes given above, and a payload SHA-256 manifest (\texttt{payloads/MANIFEST.sha256}) covering the inert synthetic substitutes shipped in place of the original malware. The release is tagged \texttt{v1.0.0} at commit \texttt{05db1fd8ea4653fb236bbac817ff32a9f3d1740e}; the archival DOI is finalized at publication.

\section{EvilModel Reimplementation}
\label{sec:evilmodel_reimpl}
The EvilModel authors did not release their code, so Experiment~3 (Section~\ref{sec:exp3}) relies on an author-developed reimplementation of the published byte-substitution method~\cite{evilmodel1,evilmodel2}. This appendix documents that reimplementation.

\subsection{Algorithm-to-Reimplementation Mapping}
\label{ssec:em_mapping}
Table~\ref{tab:em_mapping} maps each component of the published method to its counterpart in our reimplementation. The steps are read from EvilModel~2.0 (Sections~3.1--3.2, p.~6, and Section~4.1.1, p.~8~\cite{evilmodel2}) and, for the integrity header, from EvilModel~1.0~\cite{evilmodel1}. The byte-substitution write is reproduced faithfully; the carrier-selection rule is not, and that difference is analyzed in Section~\ref{ssec:em_deviations}.

\begin{table}[h]
\centering
\footnotesize
\caption{Mapping from the published EvilModel algorithm to our reimplementation. ``Faithful'' marks a component validated in Section~\ref{ssec:em_validation}; deviations are discussed in Section~\ref{ssec:em_deviations}.}
\label{tab:em_mapping}
\begin{tabular}{@{}p{0.30\columnwidth}p{0.34\columnwidth}p{0.26\columnwidth}@{}}
\hline
Original component & Our reimplementation & Fidelity \\ \hline
Carrier selection: neurons, logged for extraction & primary weight tensor of each permutable site, traversal order & Deviation \\
Half substitution ($X{=}16$): keep 2 leading bytes, overwrite low 2 & implemented & Faithful, bit-exact \\
MSB reservation ($X{=}24$): keep leading byte, overwrite low 3 & implemented & Faithful, bit-exact; excluded from Exp.~3 \\
Fast substitution & not implemented & Deviation \\
Integrity header: length and hash in a bias & 4-byte little-endian length and SHA-256 in first carrier bias & Faithful placement; digest our choice \\
Freeze-and-retrain recovery (optional) & not implemented & Deviation \\
Extraction: read header, verify & implemented & Faithful \\ \hline
\end{tabular}
\end{table}

\subsection{Validation Against Published Results}
\label{ssec:em_validation}
We validate the reimplementation against every published EvilModel result that constrains it. Eleven static checks (the worked example, the capacity ceilings, and the model volumes) pass, none fail, and one is not testable in principle (the within-parameter byte order, Section~\ref{ssec:em_byteorder}).

The first check reproduces the authors' own worked example. Table~1 of EvilModel~2.0 (p.~6~\cite{evilmodel2}) publishes the exact bit pattern each scheme produces on the parameter \texttt{0xBC40B763}. Driving our writer with the same input reproduces all four published patterns bit-for-bit: half substitution yields \texttt{0xBC40FFFF} for payload \texttt{0xFFFF} and \texttt{0xBC400000} for \texttt{0x0000}, and MSB reservation yields \texttt{0xBCFFFFFF} for \texttt{0xFFFFFF} and \texttt{0xBC000000} for \texttt{0x000000}. This validates the mask, the count of preserved bytes, and the substitution direction for both schemes.\footnote{In the MSB-reservation minimum cell the original prints the decimal $-0.031249999981374$ beside the hex \texttt{0xBCFFFFFF}, which decodes to $-0.03124999813735485$. The hex is self-consistent with the method and with the other three cells, so we match the hex and read the printed decimal as a transcription slip; the comparison is performed at the bit level.}

Three further checks confirm quantitative agreement. Running the capacity computation with every \texttt{float32} parameter as carrier reproduces both published maximum embedding rates exactly, $50.0000\%$ for half substitution and $75.0000\%$ for MSB reservation (p.~10~\cite{evilmodel2}). Model volumes computed as parameters plus buffers times four bytes agree with the published sizes within $0.06\%$: VGG16 $527.79$ against $527.87$~MB, ResNet101 $170.34$ against $170.45$~MB, and ResNet50 $97.69$ against $97.75$~MB.

A further check compares host accuracy. EvilModel embeds across the whole model, so to compare against its published ImageNet top-1 figures we ran our encoder with a whole-model carrier and evaluated over the full 50{,}000-image ILSVRC-2012 validation split with the \texttt{IMAGENET1K\_V1} weights the published baselines correspond to. Our reproduced baselines match the published Base row within $0.016$ percentage points on all three comparable architectures (MobileNetV2 $71.870$ against $71.878$, ResNet50 $76.146$ against $76.130$, ResNet101 $77.378$ against $77.374$), which confirms the data loading, label mapping, and preprocessing. Under half substitution, our accuracies agree with the published values on all seven comparable cells, with deltas from $0.000$ to $-0.038$ percentage points and ResNet101 exact. Under MSB reservation our host accuracies do not agree with the published values, a failure Section~\ref{ssec:em_deviations} traces to carrier selection rather than to the byte-level writer.

The agreement above covers the only three evaluated architectures that are comparable with EvilModel's published tables: ResNet50, ResNet101, and MobileNetV2. EvilModel's ``Vgg16'' is \texttt{vgg16\_bn} (baseline $73.360\%$), not the plain \texttt{vgg16} evaluated here (baseline $71.592\%$), and its ``Squeezenet'' is \texttt{squeezenet1\_1}. The two VGG16 volumes differ by only $0.07$~MiB, so a size check does not catch the substitution, whereas the $1.768$-point baseline-accuracy gap does.

\subsection{Implementation Details}
\label{ssec:em_impl}
The reimplementation embeds a payload into the primary weight tensors of NeuPerm's permutable sites (Section~\ref{sec:neuperm_implement}). The integrity header is 36 bytes: a 4-byte little-endian payload length followed by a 32-byte SHA-256 digest of the payload, written into the bias of the first carrier site before any payload byte. The write primitive reinterprets a \texttt{float32} tensor as \texttt{int32}, clears the low $8x$ bits of each parameter, and ORs in $x$ payload bytes, where $x{=}2$ for half substitution and $x{=}3$ for MSB reservation. Payload bytes are written across the carrier weight tensors in permutable-site traversal order. Extraction reads the 36-byte header, reads the number of payload bytes its length field gives, and verifies the SHA-256 digest. Experiment~3 uses a fill fraction of $0.25$, five seeded repeats with seeds $7$ through $11$, and torchvision \texttt{DEFAULT} pretrained weights cast to \texttt{float32}; the byte-exact and bit-error outcomes it reports are independent of the pretrained-weight variant, whereas the accuracy validation above pins \texttt{IMAGENET1K\_V1}.

\subsection{Payload-Capacity Calculations}
\label{ssec:em_capacity}
The payload capacity of the carrier is the number of its parameters times the substituted byte count, $\mathrm{capacity}=Cx$ bytes, where $C$ is the number of \texttt{float32} parameters in carrier tensors and $x$ is the substituted bytes per parameter; the integrity header is stored in a bias and consumes no payload capacity. Experiment~3 embeds a payload of $L_S=fCx$ bytes at fill fraction $f{=}0.25$. This fill fraction is not EvilModel's embedding rate. EvilModel defines $E=L_S/L_M$, the ratio of payload size to model volume $L_M=4P$ bytes for a model of $P$ \texttt{float32} parameters (Section~4.1.1, p.~8~\cite{evilmodel2}); we confirmed this definition against two published figures, $2.30/4.74=48.52\%$ and $36.9/178=20.73\%$. Substituting $L_S$ gives $E=fx(C/P)/4$. For VGG11, with $f{=}0.25$, $x{=}2$, and carrier coverage $C/P=0.9513$, this is $E=11.89\%$; across the eight evaluated architectures $E$ ranges from $2.7\%$ (MobileNetV2) to $11.9\%$ (VGG16). EvilModel's reported $48.52\%$ is an $E$ value, so our $25\%$ fill corresponds to a substantially smaller $E$.

\subsection{Byte-Order Assumptions}
\label{ssec:em_byteorder}
Our writer is little-endian within each parameter: the first payload byte becomes the least-significant byte of the substituted field, because each parameter's payload word is assembled as $\sum_j b_j \ll 8j$. This convention is independent of host endianness, because all packing is integer shift-and-mask arithmetic and the length prefix is packed little-endian, so a stego-model extracts identically on any machine. It cannot be validated against EvilModel, because its only published worked-example payloads (\texttt{0xFFFF} and \texttt{0x0000}) are byte-palindromic and so identical under either byte order; an asymmetric payload would be needed to discriminate them, and the original publishes none. The convention is thus an assumption; it is self-consistent and affects no result we report, but a payload embedded with the authors' own tool would not extract with ours if their convention is the opposite.

\subsection{Deviations From the Original Method}
\label{ssec:em_deviations}
Carrier selection is the one deviation that changes behavior. EvilModel selects carrier neurons and logs which neurons were used for extraction (Section~3.1, p.~5, and Section~4.1.1, p.~8~\cite{evilmodel2}); our reimplementation instead fills the primary weight tensors of NeuPerm's permutable sites in traversal order, which makes the corruption prefix-complete. Under MSB reservation this places host accuracy at chance, approximately $0.1\%$, at every embedding rate. Published MSB reservation gives $74.386\%$ at $E=7.53\%$, where a prefix-complete fill at that same rate measures $0.112\%$, so no single selection rule can produce both, and the original's selection is provably not prefix-complete. The published description does not specify the rule in sufficient detail to reproduce it, and its actual order remains unknown. Under half substitution, whose relative error is bounded at $2^{-7}=0.78\%$ per parameter and is survivable regardless of placement, the carrier choice is immaterial, which is why Experiment~3 is scoped to that scheme and excludes MSB reservation.

Four further deviations require only disclosure. The payload is a seeded random byte blob of known content rather than live malware, and Experiment~3 measures byte-exact recovery rather than execution. Random bytes are neither compressible nor structured, so entropy-based detection results are not comparable, and we make no detection claim. Fast substitution, the third published scheme, is not implemented. The integrity digest is SHA-256, which is our choice; the original records that a length and a hash are stored but does not fix the algorithm. EvilModel's optional freeze-and-retrain accuracy-recovery step and its trigger mechanism are also not implemented, and neither affects whether embedded bytes survive a permutation.

One architecture cannot be evaluated at all. The 36-byte integrity header must fit in the first carrier site's bias, which requires 18 elements under half substitution, whereas MobileNetV3-Small's first permutable site is a squeeze-and-excitation \texttt{fc1} whose bias has 8. MobileNetV3-Small is therefore excluded and eight architectures are evaluated, an exclusion that follows from the header placement rather than from the attack.


\bibliographystyle{elsarticle-num}
\bibliography{main}

\end{document}